\documentclass{aa}  
\usepackage{graphicx}
\usepackage{txfonts}
\usepackage{multirow}
\usepackage{color}
\usepackage{supertabular}
\usepackage{bm}
\usepackage{mathtools}
\usepackage{array}

\usepackage{hyperref}
\hypersetup{colorlinks=true, citecolor=blue, linkcolor=blue}

\setcitestyle{notesep={;}}
\begin{document} 

   \title{Revisiting the analysis of HW Vir eclipse timing data}

        \subtitle{I. A frequentist data modeling approach and a dynamical stability analysis}

\author{Ekrem. M. Esmer
        \inst{1,2}
        \and
        Özgür Baştürk\inst{2}\fnmsep
        \and Tobias C. Hinse\inst{3,4}\thanks{Corresponding author: T. C. Hinse (tchinse@gmail.com)}
        \and Selim O. Selam\inst{2}
        \and Alexandre C. M. Correia\inst{1,5}}

\institute{     CFisUC, Department of Physics, University of Coimbra, 3004-516 Coimbra, Portugal
        \and
        Department of Astronomy and Space Sciences, Ankara University, 06100 Ankara, Turkey
        \and
        Institute of Astronomy, Faculty of Physics, Astronomy and Informatics, Nicolaus Copernicus University, Grudziadzka 5, 87-100 Torun, Poland.
        \and
        Department  of  Astronomy and Space Science, Chungnam  National  University, 34134 Daejeon, Republic of Korea
        \and
        ASD, IMCCE, Observatoire de Paris, PSL Universit\'e, 77 Av. Denfert-Rochereau, 75014 Paris, France
}

\date{Received Jun 11, 2020; Accepted Feb 2, 2021}

  \abstract
   {HW Vir is a short-period binary that presents eclipse timing variations. Circumbinary planets have been proposed as a possible explanation, although the properties of the planets differ in each new study.}
   {Our aim is to perform robust model selection methods for eclipse timing variations (ETV) and error calculation techniques based on a frequentist approach for the case of the HW Vir system.}
   {We initially performed simultaneous light and radial velocity curve analysis to derive the masses of the binary. We then analyzed the eclipse timing variation of the system by fitting multiple models. To select the best model, we searched the confidence levels for the best model by creating an $\chi^2$ surface grid and bootstrap methods for each pair of parameters. We searched for stable orbital configurations for our adopted ETV model.}
   {The masses of the binary are found as $0.413 \pm 0.008\ M_{\sun}$ and $0.128 \pm 0.004\ M_{\sun}$. Under the assumption of two light time effects superimposed on a secular change, the minimum masses of the circumbinary objects are calculated as $25.0_{-2.2}^{+3.5} \ M_{Jup}$ and $13.9_{-0.45}^{+0.60}\ M_{Jup}$. The projected semi-major axes are found to be $7.8_{-1.0}^{+1.4}\ au$ and $4.56_{-0.22}^{+0.27}\ au$ in respective order. We find that this configuration is unstable within a 3$\sigma$ range on the semi-major axis and eccentricity of the outer circumbinary object.}
   {}


   \maketitle

\section{Introduction}\label{introduction}

Studies of variations in mid-eclipse timings can provide invaluable information about various  phenomena observed in eclipsing binaries such as mass exchange or loss, apsidal motion, magnetic activity, and the presence of additional bodies. In general, the analysis of eclipse timing variations (ETV) depends heavily on having a sufficient amount of precise photometric observations, which may span over many decades. Circumbinary exoplanet detections are possible due to the effect they induce on the observed eclipsing binary systems. Binaries with extremely short orbital periods are targeted due to a) their relatively low total mass that enables the detection of a potential, low-amplitude light-time effect (LiTE); and b) reducing the time spent on the retrieval of a sufficiently large number of observations. Hence, for almost all of the proposed circumbinary planets found with the ETV technique, the host binaries' orbital periods are as short as a few hours. Some relatively recent ETV studies of short-period binary systems claim detections of gravitationally bound, circumbinary objects with substellar masses. \cite{2009AJ....137.3181L} were among the first to propose circumbinary planets around HW Vir. Sinusoidal trends on eclipse timing variations for some other binaries such as NN Ser \citep{2010A&A...521L..60B}, DP Leo \citep{2010ApJ...708L..66Q}, HU Aqr \citep{2011MNRAS.414L..16Q}, NY Vir \citep{2012ApJ...745L..23Q}, Kepler-451 \citep{2015A&A...577A.146B}, Kepler-1660 \citep{2017MNRAS.468.2932G}, and GK Vir \citep{2020MNRAS.497.4022A} were also attributed to circumbinary planets. Some such cases have been re-investigated concerning their orbital stabilities (e.g., HW Vir: \cite{2012MNRAS.427.2812H}, RZ Dra: \cite{2014A&A...565A.104H}, NSVS 14256825: \cite{2013MNRAS.431.2150W}).

In this paper, we investigate trends that we observed in the ETVs of the short-period binary system HW Vir, based on our own precise photometric follow-up observations with different telescopes, space-borne observations with the Kepler space telescope, Wide Angle Survey for Exoplanets (WASP) telescopes, and timing data compiled from the published literature. We present the technical details relating to our decision-making for the quantitative model selection and perform robust error estimation based on a frequentist approach. In a follow-up paper, we plan to present results using a full Bayesian approach on a very similar dataset.

HW Vir (BD-07 3477) is an Algol-type eclipsing binary with an orbital period of 0.1167 d. The system was first identified as a subdwarf-B star by \cite{1980A&A....85..367B}, and later on \cite{1986IAUS..118..305M} revealed its binary nature. HW Vir is the prototype of binary systems consisting of a subdwarf-B (sdB) primary and a main-sequence M-type secondary dM. The system has been a topic of research since its discovery for its various aspects. \cite{1986IAUS..118..305M} made the first photometric study of the system. They found that the masses of the stars are 0.25 $M_{\sun}$ and 0.12 $M_{\sun}$ while the temperatures are 26000 K and 4700 K, respectively. From their single-lined spectrum, \cite{1996MNRAS.279.1380H} derived the temperatures as 33000 $\pm$ 800 K and 3700 $\pm$ 700 K for the binary. They did not detect any spectral signature of the secondary star due to the high contrast between the components. \cite{1999MNRAS.305..820W} were the first to claim the detection of the secondary on the spectra. They derived the semi-amplitude of the radial velocity for the secondary as $K_2$ = 275 $\pm$ 15 km/s. In order to recover the spectral signatures of the secondary, they corrected their spectra to absolute fluxes by using previous photometric observations and extracted the spectra of the secondary. \cite{2008ASPC..392..187E} calculated absolute parameters of the components based on the radial velocity measurements of the much cooler dM-component (secondary), as well as the sdB primary from additional weak absorption lines detected around the secondary eclipse due to reflections on the surface of the less massive dM-star. With these data, the masses were determined as 0.53 $\pm$ 0.08 $M_{\sun}$ and $0.15 \pm 0.03\,M_{\odot}$ by \cite{2008ASPC..392..187E}.

In a more recent study, based on Kepler (K2) data, \cite{2018MNRAS.481.2721B} showed that the sdB component displayed pulsations with frequencies up to 4600 $\mu Hz$. They also found the masses of the binary by utilizing the R\o mer effect on binaries with different masses \citep{2010ApJ...717L.108K}. Comparably to the mass derived by \cite{1986IAUS..118..305M}, they found a mass of 0.26 $M_{\sun}$ for the primary. However, the mass that they found is considerably smaller than the canonical helium core ignition mass of $\sim$0.47 $M_{\sun}$ \citep{2002MNRAS.336..449H,2009ARA&A..47..211H,2016PASP..128h2001H}, which should have taken place for the primary sdB component of HW Vir. 

The variations in the mid-eclipse timings of the system was first noticed by \cite{1994MNRAS.267..535K}. ETVs of HW Vir have been analyzed in many studies to date \citep{1999A&AS..136...27C, 2000A&A...364..199K, 2003Obs...123...31K, 2004A&A...414.1043I, 2008ApJ...689L..49Q}. \cite{1999A&AS..136...27C}, \cite{2003Obs...123...31K}, and \cite{2004A&A...414.1043I} suggested a third body for the cause of ETV within the mass ranges of a substellar object. \cite{2009AJ....137.3181L} ruled out the possible cyclic magnetic activity effect to explain the observed eclipse timings due to the absence of accompanying out-of-transit brightness variations expected by \cite{1992ApJ...385..621A} and \cite{1998MNRAS.296..893L}. \cite{2009AJ....137.3181L} explained the observed timing variations by LiTE induced by two circumbinary planets with masses $M_3 = 19.2 \ M_{Jup}$ and $M_4 = 8.5 \ M_{Jup}$. Orbital periods of these planetary mass companions were given as $P_3$ = 15.84 yr and $P_4$ = 9.08 yr with the eccentricities $e_3$ = 0.46 and $e_4$ = 0.31, respectively, in the same study. \cite{2012MNRAS.427.2812H} and \cite{2012A&A...543A.138B} questioned the validity of these orbital parameters. \cite{2012MNRAS.427.2812H} performed an orbital stability analysis of the system and reported that the orbits of the circumbinary planets would be dynamically unstable with the parameters given by \cite{2009AJ....137.3181L} on a timescale of a few thousand years. They suggested that the observed variations on mid-eclipse timings of the binary may not have been caused only by the LiTE. \cite{2012A&A...543A.138B} also found that the orbits of the planets suggested by \cite{2009AJ....137.3181L} should be unstable. They analyzed the eclipse timings of the binary system using the data set of photometric observations from 1984 to 2012 and gave a new set of parameters for the circumbinary components ($M_3$ = 14.3$\ M_{Jup}$, $M_4$ = 65$\ M_{Jup}$). They found that the orbits should be stable for more than $10^8$ yr with the configuration that they proposed.

In Section \ref{observations} of this study, we present our photometric observations of the system carried out by four different observatories. By using our photometric observations and two separate sets of radial velocity data available in the literature, we present the details of a simultaneous light and radial velocity curve analysis in Sect. \ref{lightcurveanalysis}. We formed a combination of two radial velocity data sets by \cite{1996MNRAS.279.1380H} and \cite{1999MNRAS.305..820W}. Since \cite{2008ASPC..392..187E} also published the radial velocities for the secondary star, we refer to a separate analysis of his data. After deriving the absolute parameters of the binary from simultaneous light and velocity curve analysis in Sect. \ref{lightcurveanalysis}, we present the results of an ETV analysis based on a long-baseline data set, including the recent observations in the literature as well as our own observations, which we describe in Sect. \ref{timingdatamodeling}. We attempted fits of the data with various models, extracted the statistical measures of their success, and compared them, as a result, making use of the Durbin-Watson statistics \citep{1950DurbinWatson}, $\chi^2,$ and $F$-tests. In Section \ref{erroranalysis}, we provide the details of our error analysis based on an $\chi^2$ surface search method and bootstrapping to evaluate the uncertainties on each of the model parameters. Finally, in Sect. \ref{stability}, we describe our search for dynamical stability of the orbital configuration of our adopted ETV model.

\section{Observations and data reduction}\label{observations}
Photometric observations of HW Vir, starting from 2014 Feb to 2019 Mar, were carried out with the 0.35 m telescope T35 at Ankara University Kreiken Observatory (AUKR) using $Apogee \: ALTA \: \: U47+$ CCD, 1 m telescope T100 at T\"UB\.{I}TAK National Observatory of Turkey (TUG) using the $SI \: 1100 \: Cryo, \: UV, \: AR, \: BI$ , 0.6 m telescope at Sobaeksan Optical Astronomy Observatory (SOAO) using a 2k CCD,  and the 1 m telescope at Lemmonsan Optical Astronomy Observatory (LOAO) using a 4k CCD. We acquired observations through Bessel $BVRI$ filters. The majority of observations were done through $R$ filters. All of the observations were made under clear to thin cloud conditions with various moon phases. We used 2x2 CCD binning for the observations from TUG, SOAO, and LOAO. All light curves were generated from differential photometry, and comparison stars were checked to be non-variable within nightly observation uncertainties and the limits of our observational setups. For this purpose, we used the {\sc AstroImageJ}\footnote{\url{https://www.astro.louisville.edu/software/astroimagej/}} \citep{2017ASTROIMAGEJAJ....153...77C} software package, which also allows us to correct our images for instrumental effects (bias-dark-flat corrections), perform differential aperture photometry, and detrend the airmass effect. The log of our photometric observations can be found in Table \ref{table:log_photometric_obs}.

For the LOAO observations, the technical staff encountered a software issue in recording the correct time stamp. Technical follow-up tests have subsequently identified and resolved the problem. An offset of 15 min was added to the recorded time stamp in order to obtain the correct $HJD_{UTC}$ time stamp.

The Kepler Space Telescope observed HW Vir in K2 Campaign 10 in mid-2016. We collected only short cadence photometric data for ETV modeling, since the sampling rate of the long cadence data is not sufficient to detect even the secondary eclipse at all. We calculated 1011 mid-eclipse times with {\sc Xtrema} software \citep{Bahar2015}, which makes use of Kwee-van Woerden method \citep{1956BAN....12..327K}. These data consist of 499 primary and 512 secondary mid-eclipse times.

\begin{center}
        \begin{table*}
                \begin{tabular}{ccccc}

                        \textbf{Telescope}&\textbf{Dates of Observations }&\textbf{Filters}&\textbf{Mean Phot. Errors (mag)}& \textbf{No. of Mid-Eclipse Times}\\
                        \hline \\ [-7pt]
                        &2015 Feb 21, Mar 13, Apr 17, May 15, 26; &&&\\
                        AUKR - T35&2016 Feb 3, 7, 21, 22, Mar 9, Apr 1, 27;&BVRI&9.2, 8.7, 3.5, 3.3 ($\times 10^{-3}$)&38 (23 pri; 15 sec)\\
                        &2017 Feb 2, 18, Apr 3, Jun 6, 12&&&\\
                        &2019 Mar 24&&&\\
                        TUG - T100 &2014 Feb 19, May 4; 2015 Apr 24&R&0.0020&7 (4 pri; 3 sec)\\
                        &2018 May 3&&&\\
                        SOAO &2015 Mar 21, 22; 2016 Mar 21, Apr 1, 13&R&0.0054&14 (7 pri; 7 sec)\\
                        LOAO &2016 Apr 28, 29, 30&BVR&9.2, 9.3, 17 ($\times 10^{-4}$)&7 (4 pri; 3 sec)\\ [3pt]
                        \hline\\ [-7pt]
                \end{tabular}
                \caption{Log of photometric observations with "sec" and "pri" denoting secondary and primary eclipses, respectively.}
                \label{table:log_photometric_obs}
        \end{table*}
\end{center}

\section{Simultaneous analysis of light and radial velocity curves}\label{lightcurveanalysis}

In order to derive the absolute physical parameters of the binary, we performed a simultaneous analysis of light and radial velocity curves of HW Vir. We selected our photometric observations with T35 on 2017 Feb 2 (filter $B$), 2016 Mar 9 (filter $V$), 2016 Feb 7 (filter $R$), and 2016 Apr 27  (filter $I$). The orbital phase coverage is complete and data is relatively precise (see Table \ref{table:log_photometric_obs} for mean photometric errors) compared to other observations.

The comparison and check stars for these specific observations selected for modeling were TYC 5528-596-1 and TYC 5528-655-1, which are also found to be non-variable within the observational limits. Since the observations of the four nights mentioned were selected for simultaneous light and velocity curve analysis, computations of the orbital phases and normalization were performed individually. Each light curve was normalized with respect to the mean brightness around the brightest phases of the system, which corresponds to the orbital phase between 0.43 and 0.57.

In their work, \cite{2009AJ....137.3181L} informed us of the absence of any long-term change in the light curve of HW Vir. We confirm their claim based on a comparison of the light curves spaced by two years apart from each other. Therefore, we decided to use the light curves from different observation nights in the same light curve analysis. The light curves display two eclipses with significantly different depths. This is due to the large difference between the luminosities of the stars. Other than the eclipse features, there is the reflection effect highly prominent around the second eclipse where the hot primary eclipses the cooler secondary, that is, phase 0.5. 

In order to derive system parameters through a simultaneous analysis, we collected radial velocity observations from the literature. The radial velocity data from \citet[ hereafter H96]{1996MNRAS.279.1380H} and \citet[ hereafter WS99]{1999MNRAS.305..820W} were selected for our simultaneous analysis. Both radial velocity data sets are single lined due to the high contrast in the luminosities of the binary. On the other hand, \cite{2008ASPC..392..187E} published a few radial velocity measurements for the secondary component, around the orbital phases of 0.5, by making use of the weak absorption lines that are wavelength shifted compared to the sdB's strong absorption lines. They suggested that these weak lines have formed due to the heating by reflection from the hemisphere of the secondary star facing the hot primary component. Since it was the only study so far including radial velocity data for the secondary, we decided to use the data from \citet[ hereafter ED08, see Table \ref{table:edelmann_rv_data}]{2008ASPC..392..187E} for a separate analysis. For all the simultaneous light and velocity curve analyses, we initially worked on data from WS99+H96 and adopted the results from this analysis. Then, we analyzed data from ED08 to investigate the consequences of determining the radial velocity as in \cite{2008ASPC..392..187E} on the physical parameters of the binary. We gathered the data from H. Edelmann by private communication.

\begin{center}
        \begin{table}
                \centering
                \begin{tabular}{ccc}
                        \textbf{Phase}&$\textbf{RV}_1 (\textbf{km\ s}^{-1})$&$\textbf{RV}_2 (\textbf{km\ s}^{-1})$\\
                        \hline \\ [-7pt]
                        0.05    &       -25 $\pm$3&\\
                        0.11    &       -52 $\pm$3&\\
                        0.17    &       -70 $\pm$3&\\
                        0.24    &       -81 $\pm$3&\\
                        0.30    &       -78 $\pm$3&205 $\pm$10\\
                        0.37    &       -57 $\pm$3&184 $\pm$10\\
                        0.43    &       -25 $\pm$3&126 $\pm$10\\
                        0.50    &       7 $\pm$3&\\
                        0.56    &       35 $\pm$3&-107 $\pm$10\\
                        0.63    &       63 $\pm$3&-174 $\pm$10\\
                        0.69    &       84 $\pm$3&-209 $\pm$10\\
                        0.76    &       85 $\pm$3&\\
                        0.82    &       84 $\pm$3&\\
                        0.85    &       72 $\pm$3&\\
                        0.92    &       48 $\pm$3&\\
                        0.98    &       28 $\pm$3&\\ [3pt]
                        \hline \\ [-7pt]
                \end{tabular}
                \caption{Radial velocity data from \cite{2008ASPC..392..187E}. The values in second and third column are the radial velocity measurements of the primary and the secondary companion of HW Vir.}\label{table:edelmann_rv_data}
        \end{table}
\end{center}

\begin{center}
        \begin{figure}
                \includegraphics[width=\linewidth]{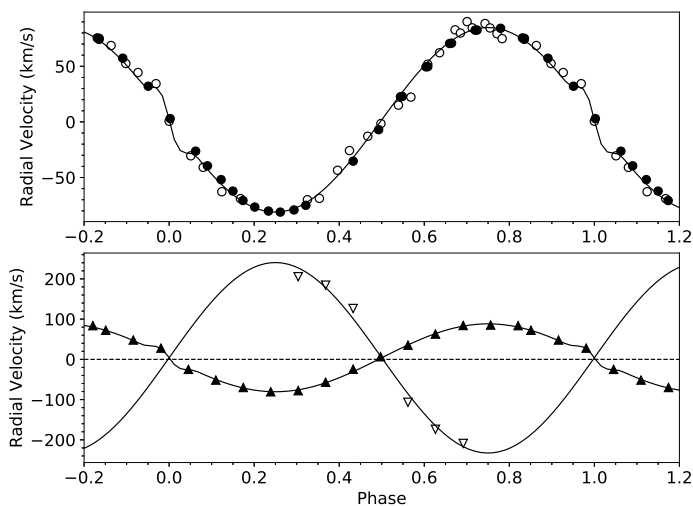}
                \caption{Radial velocity data along with synthetic curves from simultaneous light and radial velocity analysis. Combined radial velocity data of \cite{1999MNRAS.305..820W} (white) and shifted \cite{1996MNRAS.279.1380H} (black) is above. Radial velocities from \cite{2008ASPC..392..187E} is below. The white triangle marker represents the measurements for the secondary companion. See text for details.}
                \label{figure:rv_hwvir}
        \end{figure}
\end{center}

The semi-amplitudes of the radial velocities are $82.3 \pm 4$ km/s and $83.0 \pm 1.2$ km/s in WS99 and H96 datasets, respectively, which indicates that the amplitude does not change over time (agreement level of $0.17\sigma$) Radial velocity of the barycenter of the system ($V\gamma$) was given as $2.9\pm3.1$ km/s and $-9.1\pm0.9$ km/s by \cite{1999MNRAS.305..820W} and \cite{1996MNRAS.279.1380H}, respectively. Since there are about a few years of difference between the two observations of the two study, we think that such a significant (3.72$\sigma$) difference in $V\gamma$ can be the result of the uncertainties in the parameter and the lack of velocity observations of the secondary, not an actual change in the systemic velocity. The $V\gamma$ parameter does not have any impact on the binary masses. Therefore, we shifted the data of \cite{1996MNRAS.279.1380H} by the difference of $V\gamma$, to remove the velocity shift between the two datasets. We phased and combined them and used this combined set for the first simultaneous light and velocity curve analysis. The same procedure was followed for the \cite{2008ASPC..392..187E} radial velocity data forming the basis of a separate analysis. Both datasets can be seen in Fig. \ref{figure:rv_hwvir}.

We used {\sc PHOEBE} v0.31\footnote{\url{http://phoebe-project.org/1.0}}  \citep{2011ascl.soft06002P} for the simultaneous light and radial velocity curve analysis for the WS99+H96 dataset. We selected the modeling option for an unconstrained binary system. We fixed the surface temperature of the primary component ($T_{1}$) to 28500 K \citep{1999MNRAS.305..820W,2004A&A...414.1043I}, albedos for both components ($A_{1,2}$) to unity, because of the intense reflection on the secondary, gravity brightening of the primary ($g_{1}$) to unity, and that of the secondary ($g_{2}$) to 0.32. We left the remaining parameters (semi-major axis ($a$), mass ratio ($q$), systemic velocity ($V_{\gamma}$), orbital inclination ($i$), surface temperature of the secondary ($T_{2}$), surface potentials of both of the components ($\Omega_{1,2}$), luminosity ($L_{1}$), and limb darkening of the primary ($ld_{1}$) as free parameters. We assumed synchronous rotation concerning the high projected rotational velocity of the sdB component calculated as 74 $\pm$ 2 km s$^{-1}$ by \cite{2008ASPC..392..187E} as consistent with the tidal locking of the close system.

\begin{center}
        \begin{figure*}
                \centering
                \includegraphics[width=480px]{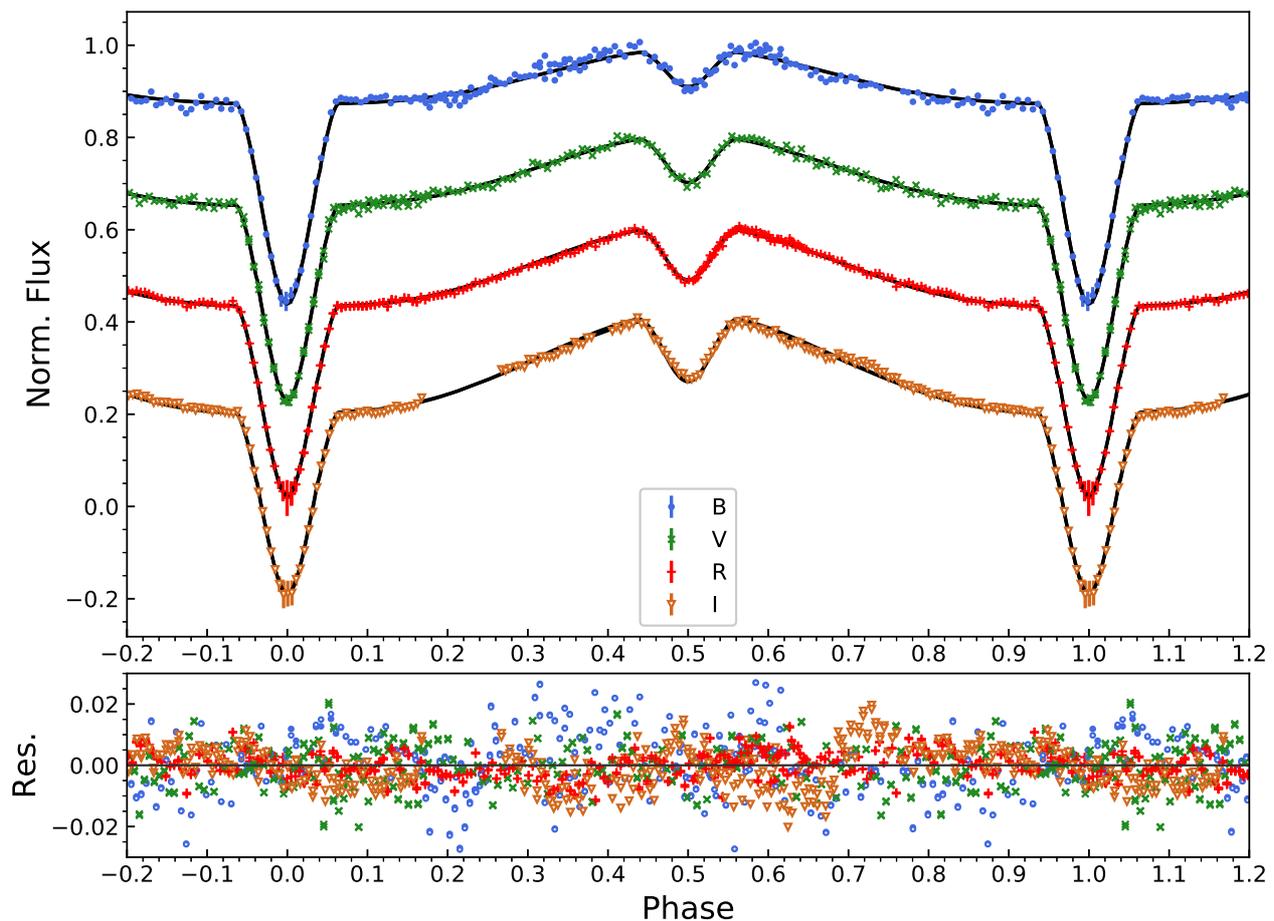}
                \caption{Observed light curves with best-fit models and their residuals in various pass-bands.}
                \label{figure:lc_curves_hwvir}
        \end{figure*}
\end{center}

Our first trial for the simultaneous analysis yielded a considerable difference for the parameters concerning the luminosities between the model curve and data especially for the $B$ filter. This difference was probably caused by the radiative properties of the hot sdB companion and the intense reflected light from the secondary companion. Therefore, we decided to derive the physical parameters of the binary using only $VRI$ filters and followed the same procedure, which resulted in an acceptable model fit and parameter values. However, there were deviations from the model in certain orbital phases. For the $V$ filter, while the model expected higher luminosities around the secondary minimum than the observed, the deviation is in the opposite direction in the $R$ band for the same phase. The model for the $I$ filter was much better than the other filters, which is most probably due to the strong reflection effect on the secondary component. To better understand the extent of the reflection effect, we analyzed each filter ($BVRI$) separately, but this time fixing the physical parameters ($T_{1,2}$, $q$, $a$, $i$, $\Omega_{1,2}$, $V_{\gamma}$) that we derived from $VRI$ solution, and adjusting the parameters concerning the luminosities ($L_{1,2}$, $ld_{1,2}$, $A_{1,2}$, $g_{1,2}$). The luminosity of the secondary ($L_{2}$) was decoupled from the temperature in this step, due to the additional luminosity coming from the reflection. The first problem we encountered in this step was the convergence of the gravity brightening of the secondary to non-physical values. Thus, we fixed $g_{2}$ to 0.32. Then, we realized that the albedo $A_{2}$ and the decoupled luminosity of the secondary $L_{2}$ were affecting the model in the same manner. Therefore, we split the analysis for each filter into two groups, one with a fixed albedo ($A_{2} = 1.0$), the other with a fixed luminosity ($L_{2}$), to the same values derived in the $VRI$ solution. The albedo values converged to nonphysical values when we fixed the luminosities. Therefore, we decided to reject these results and adopt only the results for the analyses with the fixed albedos. The resultant luminosities for the individual filters can be found in Table \ref{table:lcresults}, while the absolute parameters can be found in Table \ref{table:lc_abs_results}. Synthetic light and radial velocity curves can be seen in Fig. \ref{figure:lc_curves_hwvir} superimposed on the observational data.

We followed the same modeling approach for the ED08 dataset. We compared the models based on the level of agreement between their results as well as the absolute parameters derived from the analysis making use of the WS99+H96 and ED08 datasets by computing $ABS(X-Y)$ / $\sqrt{dX^2 + dY^2}$. Table \ref{table:lc_abs_results} clearly shows that the output absolute parameters are significantly different from each other in terms of standard deviations ($\sigma$), thus the two models strongly disagree. Radial velocity measurements of the primary for both datasets are consistent with each other. However, models for the radial velocity of the primary differ in a way that the amplitude for the model based on the ED08 dataset is slightly larger than that of WS99+H96. 

The method that \cite{2008ASPC..392..187E} followed to calculate the radial velocities of the secondary is based on the detection of the weak absorption lines from the reflected hemisphere on the secondary. The reflected hemisphere is mostly visible to the observer around phase 0.5, and the projected surface with respect to the observer decreases when phase differs. Therefore, the signal from the weak absorption lines are at maximum around 0.5. It is plausible to expect that the radial velocities of the secondary are relatively more precise and accurate around this phase, while this precision and accuracy will decrease toward phases 0.25 and 0.75. Hence, it is expected that the velocity measurements of the secondary will deviate more from the true velocities when phase diverges from 0.5. The lower panel of Fig. \ref{figure:rv_hwvir} demonstrates that the secondary velocities at phase $\sim$0.3 and phase $\sim$0.7 are below the model curve, while the velocities that are closest to phase 0.5 are above. We suspect the velocity model for the secondary has smaller amplitude than it should due to the decreased accuracy of the measurements at orbital phases, which diverge from 0.5. This explains the relatively large mass ratio found for EB08 dataset compared to WS99+H96. Thus, we interpret the strong disagreement between the parameter values as the effect of the secondary's radial velocities (only available in the ED08 dataset) on the mass ratio, which is then propagated to the remaining parameters that depend on this mass ratio.

The total mass of the binary provides key information for the minimum mass of a potential circumbinary object(s) suggested in earlier works. We decided to adopt only one of the solutions of two datasets due to the strong disagreement between the results explained above. While the measurement technique of the radial velocities of the secondary by \cite{2008ASPC..392..187E} is plausible and an interesting attempt, we suspect that the results may contain some systematic tendency making the mass ratio closer to unity. Therefore, we decided to adopt the masses of the binary components found in the analysis based on the WS99+H96 radial velocity dataset in our models for the eclipse timing variations, which we present in the next section.

\begin{center}
        \begin{table}
                \centering
                \begin{tabular}{cccc}
                        \textbf{Parameter}&\textbf{WS99+H96}&\textbf{ED08}&\textbf{Agreement}\\
                        &&&\textbf{Level ($\sigma$)}\\ [2pt]
                        \hline \\ [-7pt]
                        $i$ ($^{\circ}$)        &       81.00(7)        &       80.9(1) &0.82   \\
                        $T_1$ (K)       &       28500*  &       28500*  & -\\
                        $T_2$ (K)       &       3903(7) &       3909(9) &0.53\\
                        $\Omega_1$      &       5.02(5) &       5.04(6) &0.26\\
                        $\Omega_2$      &       2.89(2) &       3.09(2) &7.07\\
                        $A_1$   &       1*      &       1*      &-\\
                        $A_2$   &       1*      &       1*      &-\\
                        $g_1$   &       1*      &       1*      &-\\
                        $g_2$   &       0.32*   &       0.32*   &-\\
                        $q = m_2 / m_1$ &       0.310(4)        &       0.357(4)        &8.31\\[2pt]
                        \multicolumn{4}{c}{\textit{Luminosities}} \\
                        $L_1$ / $L_T$  [V]      &       0.998106(3)     &       0.998107(4)     &0.2\\
                        $L_1$ / $L_T$  [R]      &       0.996305(5)     &       0.996307(7)     &0.23\\
                        $L_1$ / $L_T$  [I]      &       0.98920(2)      &       0.9892(2)       &0\\
                        $L_2$ / $L_T$  [V]      &       0.001894        &       0.001893        &-\\
                        $L_2$ / $L_T$  [R]      &       0.003695        &       0.003693        &-\\
                        $L_2$ / $L_T$  [I]      &       0.01080 &       0.0108  &-\\ [2pt]
                        \hline \\ [-7pt]
                \end{tabular}
                \caption{Results of the simultaneous light curve and velocity curve analysis of HW Vir. Numbers in parentheses denotes the uncertainty on the last digit. The asterisk (*) denotes fixed parameter during the fitting process. $L_2$ / $L_T$ are not derived parameters ($L_2$ / $L_T$ = 1 - $L_1$ / $L_T$). See text for details.}\label{table:lcresults}
        \end{table}
\end{center}

\begin{center}
        \begin{table}
                \centering
                \begin{tabular}{cccc}
                        \textbf{Parameter}&\textbf{WS99+H96}&\textbf{ED08}&\textbf{Agreement}\\
                        &&&\textbf{Level ($\sigma$)}\\ [2pt]
                        \hline \\ [-9pt]
                        $a$ ($R_{\sun}$)&0.818(3)&0.750(3)&16.03\\
                        $M_{1}$ ($M_{\sun}$)&0.413(8)&0.307(6)&10.6\\
                        $M_{2}$ ($M_{\sun}$)&0.128(4)&0.110(3)&3.6\\
                        $R_{1}$ ($R_{\sun}$)&0.175(2)&0.161(3)&3.9\\
                        $R_{2}$ ($R_{\sun}$)&0.166(3)&0.153(3)&3.06\\
                        $L_{1}$ ($L_{\sun}$)&18.0(5)&15.3(5)&3.82\\
                        $L_{2}$ ($L_{\sun}$)&0.006(2)&0.005(2)&0.35\\
                        $M_{bol,1}$ (mag)&1.61(3)&1.79(4)&3.6\\
                        $M_{bol,2}$ (mag)&10.35(5)&10.52(5)&2.40\\
                        $\log g_{1}$ (cgs)&5.570(3)&5.512(6)&8.65\\
                        $\log g_{2}$ (cgs)&5.105(2)&5.108(2)&1.06\\ [2pt]
                        \hline \\ [-9pt]
                \end{tabular}
                \caption{Absolute parameters of HW Vir for two different radial velocity datasets. Numbers in parentheses denotes the uncertainty on the last digit. See text for details.}\label{table:lc_abs_results}
        \end{table}
\end{center}

\section{Timing data modeling}\label{timingdatamodeling}
We derived 66 mid-eclipse times in total, which consist of 38 primary and 28 secondary eclipses, from 28 nights of observations. In order to determine the mid-eclipse times from the light curves of the system, we used the Kwee-van Woerden method \citep{1956BAN....12..327K}. All mid-eclipse times were then converted from $HJD_{UTC}$ to the timescale of barycentric dynamical time $BJD_{TDB}$ \citep{2010PASP..122..935E}, which we list in Table \ref{table:minimaHWVir}.

In addition to our own observations, we collected 418 mid-eclipse times of HW Vir from the literature based on individual ground-based CCD and photoelectric observations \citep{1986IAUS..118..305M,1989IBVS.3390....1M,1991IBVS.3569....1K,1993MNRAS.261..103W,1994MNRAS.267..535K,1994IBVS.4109....1G,2000Obs...120...48K,2000IBVS.4877....1O,1999IBVS.4670....1S,1999MNRAS.305..820W,1999A&AS..136...27C,2000IBVS.4912....1A,1999IBVS.4712....1A,2003Obs...123...31K,2003IBVS.5443....1G,2009AJ....137.3181L,2000BBSAG.122....1.,2000A&A...364..199K,2002IBVS.5296....1A,2004A&A...414.1043I,2005IBVS.5643....1H,2003IBVS.5484....1A,2006IBVS.5676....1K,2005IBVS.5603....1D,2006IBVS.5741....1Z,2006VSOLJ.44,2005IBVS.5657....1H,2005IBVS.5653....1D,2006IBVS.5677....1D,2008ApJ...689L..49Q,2008IBVS.5814....1D,2007VSOLJ.45,2007OEJV...74....1B,2009IBVS.5870....1D,2012A&A...543A.138B,2009VSOLJ.48,2009IBVS.5875....1N,2009OEJV..107....1B,2009IBVS.5898....1P,2010VSOLJ.50,2011OEJV..137....1B,2011IBVS.5992....1D,2013OEJV..160....1H,2013VSOLJ.55,2014VSOLJ.56,2015IBVS.6133....1K,2014IBVS.6125....1B,2017IBVS.6196....1H,2017IBVS.6232....1K,2017OEJV..179....1J,2018MNRAS.481.2721B}. In total, 336 primary and 82 secondary minima timings have been reported within these studies. We also converted them to the $BJD_{TDB}$ time standard. Timing data derived from visual observations were not included in this study due to high uncertainties.

There are two more sources in which one can find long-term photometric observations of HW Vir, SuperWASP (SWASP) survey, and Kepler Space Telescope's K2 Campaign 10. \cite{2014A&A...566A.128L} published the mid-eclipse times of HW Vir along with some other post-common envelope binaries using SWASP data. These data include two types of timings: good times of minima (179 mid-eclipse times) and extra times (94 mid-eclipse times), as the authors name them. Good minima times have errors comparable with other modern photometric observations and almost one order of magnitude more precise compared to the extra timings. Kepler K2 data has a very short time coverage of 75 d, compared to the whole ETV dataset of $\sim$12000 d (see Fig. \ref{figure:etvdiagram_allmodels}). Therefore we used weighted average to create one point resembling all of the K2 data.

In order to draw an ETV diagram, we first corrected the linear ephemeris calculated by \cite{2012MNRAS.427.2812H} based on a linear fit to the dataset, which we provide in Eq. \ref{eq:ephemeris}.
\begin{center}
        \begin{equation}
                Min~I = BJD_{TDB}~2,450,280.28472(4) + 0.116719504(1)~E
                        \label{eq:ephemeris}
        ,\end{equation}
\end{center}
where, $Min I$ refers to the time of primary eclipse and $E$ refers to the epoch. We found the root mean square (RMS) of residuals of linear fit as 52.76 s and the reduced chi-squared $\chi_{\nu}^2$ as 89.28. The RMS of the linear fit is found to be an order of magnitude larger than the average standard errors of the observations ($\approx$ 5.6 s), which is a direct indication of an ETV variation. We formed the ETV diagram in the usual manner by subtracting the observed times of minima from that computed from the linear ephemeris that we corrected (see Fig. \ref{figure:etvdiagram_allmodels}). The resultant ETV diagram has clear signs of potentially multiple cyclic trends ans is also easily distinguishable by eye.  The amplitude of the trend appears to be changing due to the nature of the responsible mechanism(s), or there is a combination of more than one cyclic variation. We used only primary eclipse timings in the fitting procedure due to their smaller scatter over the general trend, and a potential apsidal motion is ruled out because secondary mid-eclipse timings are following the same trend with the primaries in ETV diagram. This is exactly what we expect for a circular orbit. We discarded a few of the timing data from the literature, where the published uncertainties are high, thus unreliable. A small portion of the early literature data, which follow the general trend, have mid-eclipse timing measurements without quoted uncertainties. In order to find an uncertainty estimate of these data, we calculated the mean standard errors of the data as $\sigma_{i} = 6.48 \times 10^{-5} \ d = 5.6 \ s$ and assigned this value as uncertainties.

\subsection{Analysis of the orbital period variation}\label{ls_period_analysis}

\begin{center}
        \begin{figure}
                \centering
                \includegraphics[width=\columnwidth]{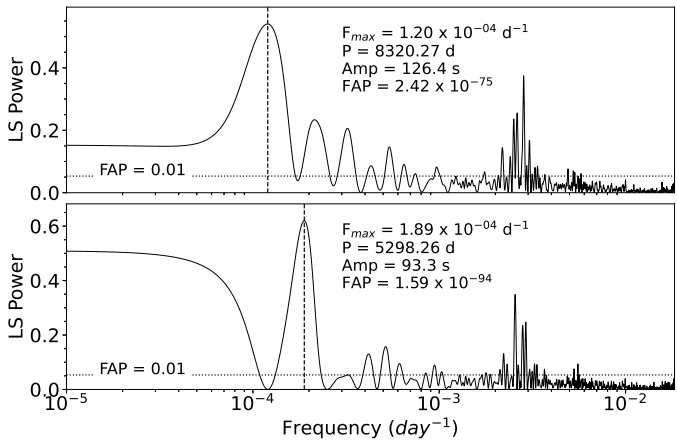}
                \caption{Lomb-Scargle spectrum of the ETV data. Upper panel is for the whole dataset, and the lower panel is for the residuals from the highest frequency of the whole dataset. We plot the 1\% FAP line for reference.}
                \label{figure:ls_freq_spec}
        \end{figure}
\end{center}

Since the mid-eclipse timings data form an unevenly distributed dataset, we made use of a Lomb-Scargle (LS) periodogram which we computed with our own {\sc PYTHON} code based on the {\sc Astropy}\footnote{http://www.astropy.org} package \citep{astropy:2013,astropy:2018} functions to detect the periodicities. We found the highest peak in the period spectrum at $f_1 = 1.20 \times 10^{-4}\,\rm d^{-1}$ with an amplitude of 126.4 s and false alarm probability (FAP) of $2.4 \times 10^{-75}$. There are some other peaks around $2.7 \times 10^{-3}\,\rm d^{-1}$, which are due to the annual sampling of the ETV data. We then removed the highest frequency from the original dataset and found a second frequency at $f_2 = 1.89 \times 10^{-4}\,\rm d^{-1}$ with an amplitude of 93.3 s and FAP of $1.6 \times 10^{-94}$ on the residual spectrum (see Fig. \ref{figure:ls_freq_spec}). Since we only removed $f_1$ from the ETV data, the frequencies arose from the annual sampling remained in the second spectrum as well. 

We also removed $f2$ from the ETV dataset to check any remaining periodicity. The highest peak on the residual spectrum corresponds to a period more than 100 times longer than the total time span of the dataset. This is consistent with a period decay, a downward parabola in the ETV, or a very long period sinusoidal. The amplitude of this frequency is almost 2 d, which is unlikely to be a part of a long period sinusoidal, since the orbital period of the binary is only 2.8 h. Therefore, we interpret this signal as a secular change.

In order to avoid overfitting the data and for simplicity, we decided to stick to a maximum of two cyclic models in modeling the ETV. Therefore, we investigated the potential of the first two peaks in our frequency analysis with LS periodograms, found at $f_1$ and $f_2$ corresponding to periods of 8320 d ($\sim$22.8 yr) and 5298 d ($\sim$14.5 yr), respectively. Considering the total time span of the data ($\sim$13200 d), it should cover more than one full cycle of the variation with $f_1$ and more than two cycles with $f_2$. As a result, we decided to use these first two peaks in a first attempt to model the observed ETV.

\subsection{ETV modeling}\label{etvmodeling}

In order to account for the accumulation of the uncertainties on the ephemeris parameters, we added a linear expression to each of the models that we think have the potential to represent the observed ETV trend.
The lowest significant frequency found in the LS periodogram could be interpreted as a secular change of the eclipse timings. Therefore, we decided to select the models to allow for a quadratic term, $\beta$, to be added if needed. Therefore ETV models ($T_C$) with $\beta$ would have a form of

\begin{equation}
                T_C = (T_0+\Delta T_{0})+(P_0 + \Delta P_{0})E+ \beta E^2 +\tau(E)
.\end{equation}

Here, $\Delta T_0$ and $\Delta P_0$ are the correction terms on the reference mid-eclipse time ($T_0$) and the orbital period of the binary ($P_0$), respectively, and $\tau(E)$ is a LiTE model(s). The models without the quadratic term will have a form similar to the one above, but without the $\beta E^2$ term. By using two different groups of models with and without a quadratic term, it can be tested if a secular change will improve the fits based on the statistical significance of the results.

\begin{center}
        \begin{figure*}
                \centering
                \includegraphics[width=520px]{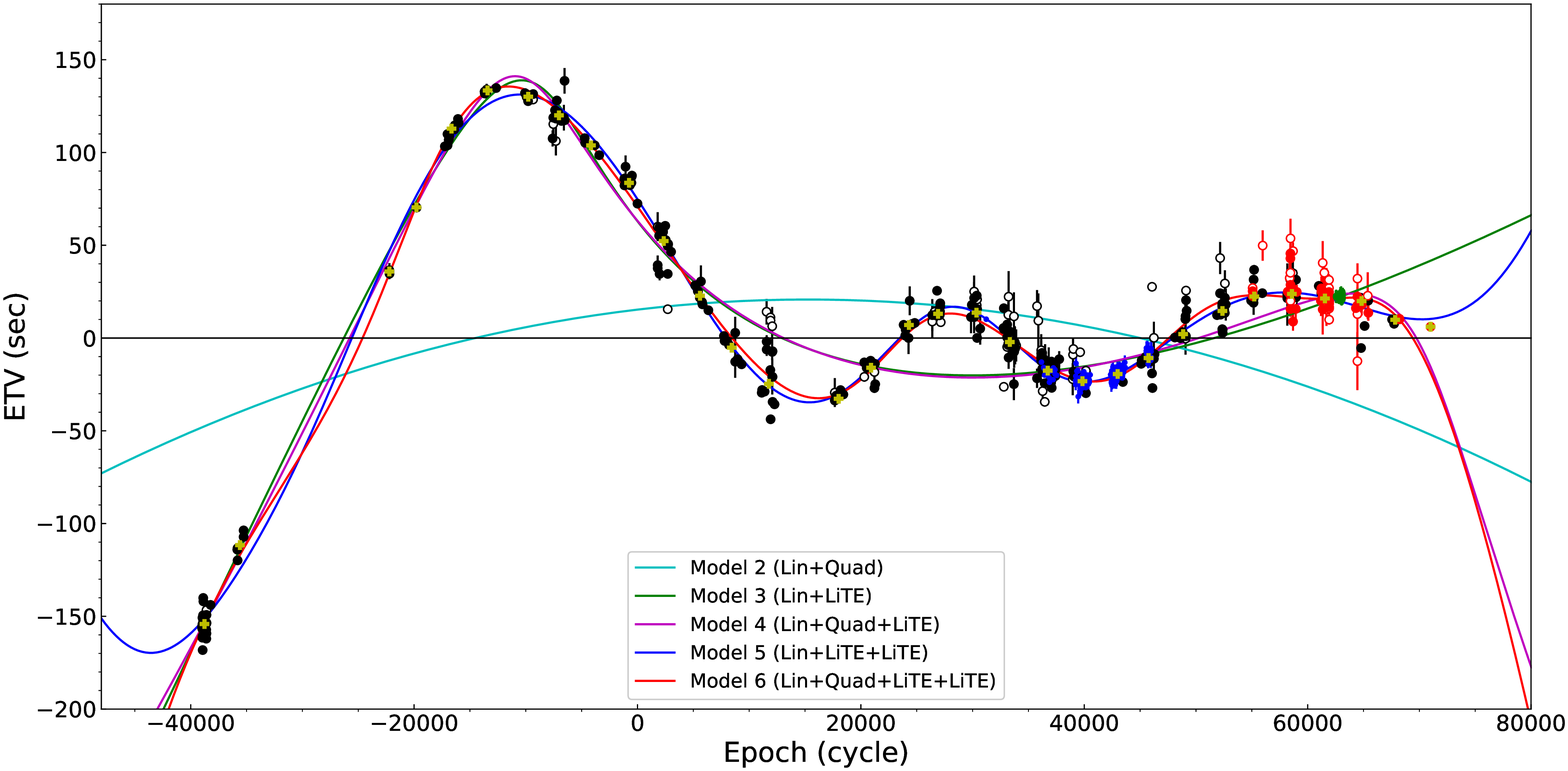}
                \caption{ETV diagram of HW Vir with best-fit curves of all models. The ETV diagram is corrected by the results of the linear fit. Color-filled circles represents the primary mid-eclipse timings, while white-filled ones represent the secondaries. The black markers are for data from the literature, blue ones are for SWASP, green for Kepler K2, red for our own observations, and yellow markers represent seasonal binned data.}
                \label{figure:etvdiagram_allmodels}
        \end{figure*}
\end{center}

From a visual inspection of the phased light curve (Fig. \ref{figure:lc_curves_hwvir}) and the ETV diagram (Fig. \ref{figure:etvdiagram_allmodels}), we do not see any evidence for apsidal motion, so we discarded it. The signals found in LS analysis may be due to LiTE caused by additional bodies or due to magnetic activity. LiTE would lead to a periodic variation with constant amplitude, while magnetic activity causes cyclic changes with variable amplitudes both in the short term (modulation due to stellar spots with rotation) and in the long run (magnetic activity cycle). We fit a sinusoidal model without the eccentricity term with the assumption of magnetic activity modulation of a secondary star as described in \cite{1992ApJ...385..621A}. The corresponding period is 77 yr, and the magnetic field should be $B \approx 39 \ kG$, which is an unrealistic level of activity expected from an M-type dwarf. Therefore, we decided not to assume magnetic-activity-induced, cyclic orbital period variations in our models. For the sake of simplicity, we used the analytical expression of LiTE \citep{1959AJ.....64..149I} for an initial model of the cyclic trends and discuss the choice between the mechanisms in Sect. \ref{conclusion}.

The models that we used to fit the ETV data are, linear (Model 1), linear + quadratic (Model 2), linear + LiTE (Model 3), linear + quadratic + LiTE (Model 4), linear + two LiTE (Model 5) and linear + quadratic + two LiTE (Model 6). The analytical expressions for each component of the models and the short-hand notations are shown in Table \ref{table:analyticexpressiontable}. These models are fit separately, and corresponding goodness-of-fit statistics (RMS, $\chi^2$, $\chi_{\nu}^2$) are calculated for each of them. For all the models that include LiTE, the mass function for a circumbinary object can be calculated as described in, for example, \cite{2018A&A...620A..72W}:
\begin{center}
        \begin{equation}
        f(m_3) = \frac{M_3^3 \sin^3 i_3}{(M_{bin} + M_3)^2} = \Bigg[ \frac{173.15 A}{\sqrt[]{1 - e_3^2 \cos^2 \omega_3}} \Bigg]^3 \frac{1}{P_3^2}
        .\end{equation}
\end{center}
Here, $M_3$ is the mass, $i_3$ is the orbital inclination, $e_3$ is the eccentricity, $P_3$ is the orbital period, $\omega_3$ is the argument of periastron of the circumbinary object, and $A$ is the semi-amplitude of LiTE, while $M_{bin}$ is the total mass of the binary stars. $P_3$ is in units of years, masses are in units of solar masses, and the constant 173.15 arises from the conversion of units. Assuming the stellar binary as a single object with the total mass of the binary, the minimum mass of the circumbinary object, $M \sin \! i$, regarding each LiTE can be found using an iterative method if the mass of the binary is already known. For the case of HW Vir, we used the masses that we derived from simultaneous light and radial velocity curve analysis in Sect. \ref{lightcurveanalysis}. 
 
\begin{center}
        \begin{table}
                \centering
                \begin{tabular}{cc}
                        \textbf{Model Component}&\textbf{Analytic Expression}\\ [2pt]
                        \hline \\ [-6pt]
                        Linear (Lin)&$\Delta P E + \Delta T_{ref}$\\
                        Quadratic (Quad)&$\beta E^2$\\
                        Light-time effect$^1$ (LiTE)& $\frac{A}{\sqrt[]{1-e^2\cos^2\omega}}$\\ 
                        &$\times~\big[\frac{1-e^2}{1+e\cos v}\sin(v+\omega)+e\sin\omega\big]$\\
                        \\
                        \textbf{Model (No. of param.)}&\textbf{Short Hand Notation}\\  [2pt] \hline \\ [-6pt]
                        Lin (2)&Model 1\\
                        Lin+Quad (3)&Model 2\\
                        Lin+LiTE (7)&Model 3\\
                        Lin+Quad+LiTE (8)&Model 4\\
                        Lin+ 2$\times$LiTE (12)&Model 5\\
                        Lin+Quad+ 2$\times$LiTE (13)&Model 6\\ [2pt]
                        \hline \\ [-6pt]
                \end{tabular}
                \caption{Analytical expressions of the model components and model list. We note that short-hand notations are ordered by the increased number of parameters. ($^1$ \citealt{1959AJ.....64..149I})}
                \label{table:analyticexpressiontable}
        \end{table}
\end{center}

We performed the fitting procedure on both the original dataset and the mid-eclipse times binned to remove the seasonal variations. Since the cyclic trends in the ETV of HW Vir that we found in frequency analysis are of the order of tens of years, seasonal averaging of one year should not diminish the periodic signals and is therefore plausible. By using seasonal averaged data, we minimized the effect of the outliers on the parameter values and their uncertainties, which made it possible to compare with the results of the fitting to the original dataset.

We used our own {\sc PYTHON} code that makes use of the Levenberg-Marquardt algorithm for a nonlinear least-squares fitting (\citealt{Levenberg.1944}, \citealt{marquardt:1963}) based on the {\sc LMFIT} package functions by \citealt{newville_2014_11813}) for the fitting procedure. Using initial parameter values, the Levenberg-Marquardt algorithm can find local minima in the parameter space. Since the Levenberg-Marquardt algorithm requires initial parameter guesses, we used the amplitude and period values derived from a Lomb-Scargle periodogram to derive initial guess values for the models including LiTE. Initial values for the remaining parameters were found from several trial test-fits and a visual inspection as implemented in a spreadsheet program. The results containing the best-fit parameter values, RMS, and $\chi_{\nu}^2$ for every model fit to the whole range of data, as well as seasonal binned data, can be found in Table \ref{table:OCresultsHWVir}. Best-fit curves for each model fit can be seen in Fig. \ref{figure:etvdiagram_allmodels} overplotted on the full dataset and the averaged data. In the following, we report the results of model fits to the original dataset and to the seasonal binned dataset. 

Model 1 is the linear model that we used to update the ephemeris information as we mentioned earlier. RMS and $\chi_{\nu}^2$ for Models 1 and 2, for fitting original and seasonal binned datasets are large enough to indicate the need for further improvement to the models. The difference between the goodness-of-fit statistics for two datasets arises from $i)$ the difference of the model parameters, and $ii)$ the difference of $\nu$ and standard errors of data points between two datasets.

With the addition of LiTE to Model 3, fit statistics decreased dramatically compared to Models 1 and 2. However, the best-fit results look far from convincing. For the original dataset, the model fit indicates an additional body with a very wide, long period ($10^6 d \approx 2750 yr$) and extremely eccentric orbit, which coincidentally should have passed from its superior conjunction relative to Earth within the time frame of the dataset ($\sim$35 yr). The results for a seasonal binned dataset also indicate a similar, highly unlikely configuration with a relatively longer period.

The addition of a quadratic term to Model 4 resulted in almost the same fit statistics. The orbital period was found to be 9399 d (26 yr) for the original dataset and 10499 d (29 yr) for binned dataset with almost the same eccentricity ($e \sim 0.5$). While the best-fit parameters are in an acceptable range, the fit statistics imply that Model 4 is not a plausible explanation to the ETV.

We introduced a second LiTE with Model 5, for which the fit statistics were found to be within acceptable limits, and in fact the best of all of the models. $\chi_{\nu}^2$ for the original data is around unity, and RMS is $\sim$5.6 s. However, periods, ETV amplitudes ($A_{LiTE}$), eccentricities and corresponding minimum masses ($M\sin\!i$), and semi-major axes ($a\sin\!i$) calculated for the original dataset are almost identical for each of the two LiTE models included in Model 5. This infers that there should be two objects with almost identical orbital and physical properties according to the best-fit solution of Model 5. The only considerable difference is the argument of periastron, $\omega$, and time of periastron passage, $T_{0,LiTE}$. $M\sin\!i$ values indicate that the masses of the additional bodies are at the limit of stellar masses, and the uncertainties of $M\sin\!i$ and $a\sin\!i$ are enormous. The minimum masses calculated from the best-fit parameters values for binned dataset are even higher than the mass of the secondary companion of the binary. We highly doubt this solution represents the case for HW Vir.

\begin{center}
        \begin{figure*}
                \centering
                \includegraphics[width=520px]{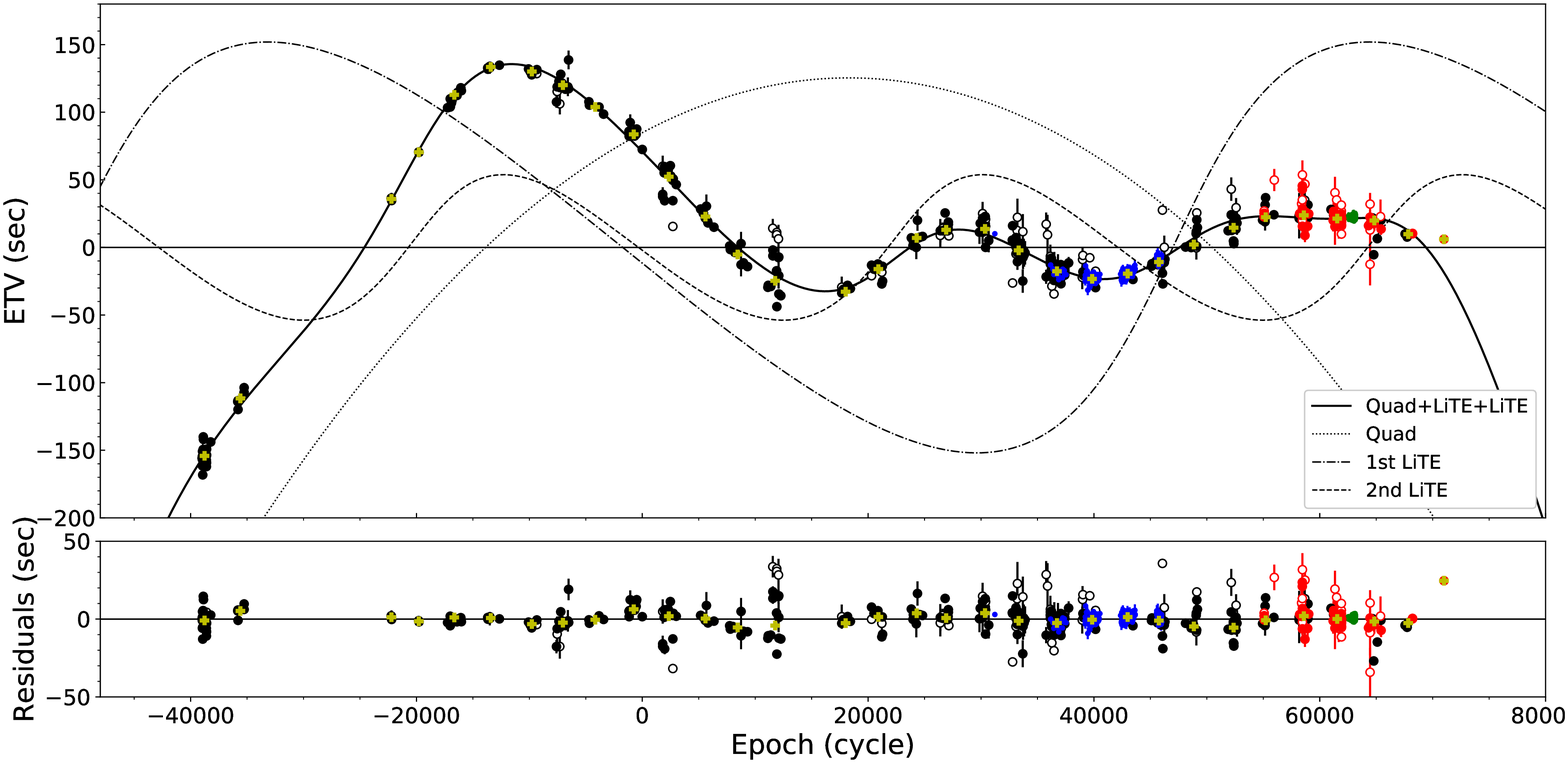}
                \caption{ETV diagram of HW Vir with best-fit curve of Model 6 and its residuals. Color-filled circles represent the primary mid-eclipse timings, while white-filled ones represents the secondaries. The black markers represent data from the literature, blue ones are for SWASP, green for Kepler K2, red for our own observations, and yellow markers represent seasonal binned data.}
                \label{figure:etvdiagram_model6}
        \end{figure*}
\end{center}

In terms of fit statistics and best-fit parameter values, one of the satisfactory ETV models has been achieved for the case of Model 6 (Fig. \ref{figure:etvdiagram_allmodels} and Fig. \ref{figure:etvdiagram_model6}). Both the fits for original and seasonal binned dataset gives almost the same solution. LiTE results of Model 6 correspond to $M\sin\!i$ of 24.6 and 13.89 $M_{Jup}$, respectively. While the periods of Model 6 differ from the results of LS analysis, they still fall in a time span shorter than that of the dataset. The orbits are eccentric and seem to be co-oriented in space as $\omega$ values are almost the same.

The best-fit values for the binned dataset are almost the same as original dataset. The orbits are slightly closer to each other, and the outer orbit is somewhat more eccentric compared to the best-fit values of the original dataset. $\chi_{\nu}^2$ statistics of the binned dataset indicate overfitting or overestimated uncertainties. As \citet[p.~107]{Hughes2010} mentioned, the number of data points ($N$) has to be similar to $\nu$ in order to have an $\chi_{\nu}^2$ of unity. However, for seasonal binned data, the ratio $N$ / $\nu$ $\approx$ 1.68, while the same ratio for the original dataset is 1.03. Therefore, the $\chi_{\nu}^2$ statistic for the binned dataset may not be a suitable goodness-of-fit indicator.

The radial velocity semi-amplitudes corresponding the two LiTEs of the Model 6 solution are both $\sim 0.2 kms^{-1}$ with the assumption of $i = 90^\circ$. Therefore, we do not expect a relatively small semi-amplitude to be detectable from the radial velocity datasets that we used in Sect. \ref{lightcurveanalysis} in terms of their observational uncertainties, the number of observations, and that the system is a single lined eclipsing binary.

Finally, we performed a frequency analysis on the residuals of Model 6. The LS periodogram of the residuals from the original dataset has its highest peak at $f_{res} = 5.9 \times 10^4$ d$^{-1}$, which corresponds to a period of 1690.7 d and a small FAP value of $6 \times 10^{-4}$. However, the amplitude of the peak is 3.97 s, which is less than the mean standard deviation ($\sim$5.6 s) of the timing data and less than the RMS for Model 6 fit to the original data ($\sim$5.7). We did not find any similar frequency with a FAP less than 0.1 on the LS periodogram of the residuals of the binned dataset. Therefore, we interpret $f_{res}$ to resemble the sampling frequency of the original dataset or its multiples.

\begin{center}
        \begin{table*}
                \centering
                \begin{tabular}{cccccccc}
                        &\multicolumn{1}{c}{\textbf{Parameter}} & \textbf{Unit}& \multicolumn{2}{c}{\textbf{Original Dataset}} & \multicolumn{2}{c}{\textbf{Seasonal Binned Dataset}}  \\ [2pt] \hline \\ [-9pt]
                        \multirow{4}{*}{\begin{sideways} \textbf{Model 1}~ \end{sideways}}
                        &$\chi_{\nu}^2$ && \multicolumn{2}{c}{$89.28$} & \multicolumn{2}{c}{$125.43$}\\
                        &RMS \: &s & \multicolumn{2}{c}{$52.764$} & \multicolumn{2}{c}{$60.674$}\\
                        &$T_0$ \: & $BJD_{TDB}$ & \multicolumn{2}{c}{$2,450,280.28471 \pm 4\times \: 10^{-5}$} & \multicolumn{2}{c}{$ 2,450,280.2850 \pm 2 \times 10^{-4}$} \\
                        &$P_0$ \: & $d$ & \multicolumn{2}{c}{$0.116719504 \pm 1\times \: 10^{-9} $} & \multicolumn{2}{c}{$0.116719502 \pm 4 \times 10^{-9} $} \\[2pt]
                        \hline \\ [-9pt]
                        
                        \multirow{5}{*}{\begin{sideways} \textbf{Model 2}~ \end{sideways}}
                        
                        &$\chi_{\nu}^2$ && \multicolumn{2}{c}{$79.5
                                2$} & \multicolumn{2}{c}{$118.26$}\\
                        &RMS \: &s & \multicolumn{2}{c}{$49.747
                                $} & \multicolumn{2}{c}{$57.924$}\\
                        &$T_0$ \: & $BJD_{TDB}$ & \multicolumn{2}{c}{$ 2,450,280.28489\pm 5 \times 10^{-5} $} & \multicolumn{2}{c}{$ 2,450,280.28513 \pm 2 \times \: 10^{-4} $} \\
                        &$P_0$ \: & $d$ & \multicolumn{2}{c}{$ 0.116719513\pm 1\times \: 10^{-9} $} & \multicolumn{2}{c}{$ 0.116719511 \pm 6 \times \: 10^{-9}$} \\
                        &$\beta$ \: &$d \ cycle^{-2}$ & \multicolumn{2}{c}{$-2.71 \times 10^{-13} \pm 3 \times 10^{-14} $} & \multicolumn{2}{c}{$-2.3 \times 10^{-13} \pm 1 \times 10^{-13} $}\\
                        \hline \\ [-9pt]
                        
                        \multirow{12}{*}{\begin{sideways} \textbf{Model 3}~ \end{sideways}}
                        
                        &$\chi_{\nu}^2$ && \multicolumn{2}{c}{$5.58$} & \multicolumn{2}{c}{$9.34$}\\
                        &RMS \: &s & \multicolumn{2}{c}{$13.126$} & \multicolumn{2}{c}{$15.114$}\\
                        &$T_0$ \: & $BJD_{TDB}$ & \multicolumn{2}{c}{$ 2,450,280.3 \pm 0.1 $} & \multicolumn{2}{c}{$2,450,280.3 \pm 0.1 $} \\
                        &$P_0$ \: & $d$ & \multicolumn{2}{c}{$ 0.11671961 \pm 3 \times 10^{-8} $} & \multicolumn{2}{c}{$ 0.11671959 \pm 1 \times 10^{-8} $} \\
                        &$T_{0,LiTE}$ \: &$BJD_{TDB}$ &  \multicolumn{2}{c}{$ 2,449,352 \pm 104 $}  &  \multicolumn{2}{c}{$ 3,595,768 \pm 8 \times 10^{6} $}  \\
                        &$P_{LiTE}$ \: &$d$ &  \multicolumn{2}{c}{$ 1.0 \times 10^6 \pm 7 \times 10^6 $}  &  \multicolumn{2}{c}{$ 1.2 \times 10^{6} \pm 8 \times 10^{6} $}\\
                        &$e$ \: &-&  \multicolumn{2}{c}{$ 0.98 \pm 0.1 $}  &  \multicolumn{2}{c}{$ 0.98 \pm 0.1 $}\\
                        &$A_{LiTE}$ \: &$s$ &  \multicolumn{2}{c}{$ 2530 \pm 1.2 \times 10^4 $}  &  \multicolumn{2}{c}{$2527 \pm 1 \times 10^{4} $}  \\
                        &$\omega$ \: &$^{\circ}$ &  \multicolumn{2}{c}{$ 133 \pm 7 $}  &  \multicolumn{2}{c}{$129 \pm 11$}  \\
                        &$f(m)$\:&$M_{Jup}$ &  \multicolumn{2}{c}{$0.04 \pm 0.9$}  &  \multicolumn{2}{c}{$ 0.03 \pm 0.6$}  \\
                        &$M\sin\!i$\:&$M_{Jup}$ &  \multicolumn{2}{c}{$25 \pm 170$}  &  \multicolumn{2}{c}{$ 21 \pm 140$}  \\
                        &$a\sin\!i$\:&$au$ &  \multicolumn{2}{c}{$156 \pm 1300$}  &  \multicolumn{2}{c}{$ 170 \pm 1400 $}  \\
                        \hline \\ [-9pt]
                        
                        \multirow{13}{*}{\begin{sideways} \textbf{Model 4}~ \end{sideways}}
                        &$\chi_{\nu}^2$ && \multicolumn{2}{c}{$5.26$} & \multicolumn{2}{c}{$6.69$}\\
                        &RMS \: &s & \multicolumn{2}{c}{$12.726$} & \multicolumn{2}{c}{$12.533$}\\
                        &$T_0$ \: & $BJD_{TDB}$ & \multicolumn{2}{c}{$ 2,450,280.28554 \pm 2 \times 10^{-5} $} & \multicolumn{2}{c}{$2,450,280.28546 \pm 8 \times 10^{-5} $} \\
                        &$P_0$ \: & $d$ & \multicolumn{2}{c}{$ 0.1167195268 \pm 8 \times 10^{-10}$} & \multicolumn{2}{c}{$0.11671954 \pm 1 \times 10^{-8}$} \\
                        &$\beta$ \: &$d \ cycle^{-2}$ & \multicolumn{2}{c}{$-7.21 \times 10^{-13} \pm 1 \times 10^{-14} $} & \multicolumn{2}{c}{$-7.76 \times 10^{-13} \pm 7 \times 10^{-14} $}\\
                        &$T_{0,LiTE}$ \: &$BJD_{TDB}$ &  \multicolumn{2}{c}{$ 2,449,127 \pm 51$}  &  \multicolumn{2}{c}{$2,449,662 \pm 281$}  \\
                        &$P_{LiTE}$ \: &$d$ &  \multicolumn{2}{c}{$ 9399 \pm 103$}  &  \multicolumn{2}{c}{$ 10499 \pm 1708 $}\\
                        &$e$ \: &-&  \multicolumn{2}{c}{$ 0.52 \pm 0.03 $}  &  \multicolumn{2}{c}{$ 0.51 \pm 0.08$}\\
                        &$A_{LiTE}$ \: &$s$ &  \multicolumn{2}{c}{$ 101 \pm 2$}  &  \multicolumn{2}{c}{$107 \pm 14$}  \\
                        &$\omega$ \: &$^{\circ}$ &  \multicolumn{2}{c}{$115 \pm 2$}  &  \multicolumn{2}{c}{$145 \pm 15$}  \\
                        &$f(m)$\:&$M_{Jup}$ &  \multicolumn{2}{c}{$0.0140 \pm 0.0008$}  &  \multicolumn{2}{c}{$ 0.017 \pm 0.09$}  \\
                        &$M\sin\!i$\:&$M_{Jup}$ &  \multicolumn{2}{c}{$16.8 \pm 0.4$}  &  \multicolumn{2}{c}{$ 17.8 \pm 3.2$}  \\
                        &$a\sin\!i$\:&$au$ &  \multicolumn{2}{c}{$7.0 \pm 0.2$}  &  \multicolumn{2}{c}{$7.5 \pm 1.7 $}  \\
                        \hline \\ [-9pt]
                        
                        \multirow{12}{*}{\begin{sideways} \textbf{Model 5}~ \end{sideways}}
                        
                        &$\chi_{\nu}^2$ && \multicolumn{2}{c}{$1.04$} & \multicolumn{2}{c}{$0.68$}\\
                        &RMS \: &s & \multicolumn{2}{c}{$5.625$} & \multicolumn{2}{c}{$3.640$}\\
                        &$T_0$ \: & $BJD_{TDB}$ & \multicolumn{2}{c}{$2,450,280.28456 \pm 4 \times 10^{-5} $} & \multicolumn{2}{c}{$2,450,280.28458 \pm 6 \times 10^{-5} $} \\
                        &$P_0$ \: & $d$ & \multicolumn{2}{c}{$ 0.1167195065 \pm 4 \times 10^{-10} $} & \multicolumn{2}{c}{$0.1167195059 \pm 8 \times 10^{-10}$} \\
                        &$T_{0,LiTE}$ \: &$BJD_{TDB}$ & $ 2,446,408 \pm 448 $ & $ 2,453,385 \pm 670 $       & $ 2,446,482 \pm 1885 $& $ 2,453,621 \pm 3199 $\\
                        &$P_{LiTE}$ \: &$d$ & $ 7800 \pm 253 $ & $ 7448 \pm 369 $                           & $ 7645 \pm 1200 $& $ 7442 \pm 1368 $\\
                        &$e$ \: &-& $ 0.19 \pm 0.04 $ & $ 0.18 \pm 0.02 $                                                         & $ 0.18 \pm 0.2 $& $ 0.17 \pm 0.02$\\
                        &$A_{LiTE}$ \: &$s$ & $ 480 \pm 801$ & $ 467 \pm 801 $                                   & $ 786 \pm 9761 $& $ 773 \pm 9761 $\\
                        &$\omega$ \: &$^{\circ}$ & $ 242 \pm 36 $ & $ 20 \pm 34 $                                                        & $ 239 \pm 158 $& $ 33 \pm 166$\\
                        &$f(m)$\:&$M_{Jup}$ & $ 2 \pm 10 $ & $ 2 \pm 11 $                                                               & $ 10 \pm 350 $& $ 10 \pm 400 $\\
                        &$M\sin\!i$\:&$M_{Jup}$ & $ 97 \pm 180 $ & $ 98 \pm 190 $                                   & $ 173 \pm 250 $& $ 175 \pm 260 $\\
                        &$a\sin\!i$\:&$au$ & $6 \pm 14 $ & $5 \pm 14 $                                                          & $ 5 \pm 100 $& $ 5 \pm 100 $\\
                        \hline \\ [-9pt]
                        
                        \multirow{13}{*}{\begin{sideways} \textbf{Model 6}~ \end{sideways}}
                        
                        &$\chi_{\nu}^2$ && \multicolumn{2}{c}{$1.07
                                $} & \multicolumn{2}{c}{$0.68$}\\
                        &RMS \: &s & \multicolumn{2}{c}{$5.699$} & \multicolumn{2}{c}{$3.556$}\\
                        &$T_0$ \: & $BJD_{TDB}$ & \multicolumn{2}{c}{$2,450,280.28569 \pm 8 \times 10^{-5}$} & \multicolumn{2}{c}{$ 2,450,280.2855 \pm 1 \times 10^{-4}$}\\
                        &$P_0$ \: & $d$ & \multicolumn{2}{c}{$0.116719556 \pm 3 \times 10^{-9}$} & \multicolumn{2}{c}{$ 0.116719549 \pm 5 \times 10^{-9}$}\\
                        &$\beta$ \: &$d \ cycle^{-2}$ & \multicolumn{2}{c}{$-1.40\: \times \: 10^{-12} \ \pm \ 9\ \times\ 10^{-14} $} & \multicolumn{2}{c}{$-1.17\times 10^{-12} \pm 1 \times 10^{-13}$}\\
                        &$T_{0,LiTE}$ \: &$BJD_{TDB}$ & $2,444,369 \pm 305 $ & $2,452,958 \pm 68 $         & $2,444,744 \pm 490$ & $ 2,453,024 \pm 166$\\
                        &$P_{LiTE}$ \: &$d$ & $11391 \pm\ 299$ & $4958 \pm\ 30 $                    & $11006 \pm 478 $ & $ 5040 \pm 58$\\
                        &$e$ \: &-& $0.45 \pm 0.01$ & $0.27 \pm 0.02$                                                   & $0.50 \pm 0.03$ & $ 0.25 \pm 0.05 $\\
                        &$A_{LiTE}$ \: &$s$ & $152 \pm 9$ & $54 \pm 1 $                                                 & $133 \pm 13$ & $55 \pm 2 $\\
                        &$\omega$ \: &$^{\circ}$ & $0 \pm\ 2 $ & $13 \pm 5 $                                                     & $358 \pm 4$ & $13 \pm 13$\\
                        &$f(m)$\:&$M_{Jup}$ & $0.043 \pm 0.008$ & $ 0.0079 \pm 0.0005 $            & $0.034 \pm 0.010 $ & $ 0.0082 \pm 0.0011 $\\
                        &$M\sin\!i$\:&$M_{Jup}$ & $24.6 \pm 1.5 $ & $ 13.89 \pm 0.33 $                      & $22.8 \pm 2.4 $ & $14.0 \pm 0.6$\\
                        &$a\sin\!i$\:&$au$ & $7.8 \pm 0.7 $ & $4.56 \pm 0.16 $                                       & $7.7 \pm 1.1 $ & $4.61 \pm 0.30$\\
                        \hline \\ [-9pt]
                \end{tabular}
                \caption{Parameter values and formal uncertainties for the ETV models of HW Vir.}\label{table:OCresultsHWVir}
        \end{table*}
\end{center}

\subsection{Durbin-Watson test}\label{dwtest}

Often within the framework of a least-squares minimization, the residuals are qualitatively assessed to judge any trends or structures (auto-correlations) that might be present in the data. The nature of a structure is either astrophysical, and/or a non-Gaussian measurement process is involved. If the measurement errors are assumed to follow a Gaussian distribution with $\sim$$N(0, \sigma^2$), the presence of auto-correlation (change of variance with time) could be interpreted as an additional signature of astrophysical origin. We applied the Durbin-Watson ($DW$) statistic \citep{Hughes2010} to quantitatively assess the quality of a given model fit. In addition, we drew so-called lag-plots to visualize the underlying concept of the $DW$ statistic for each residual set. A lag plot is constructed by plotting the residuals of a given model against a lagged residual by selecting a lag interval $k = 1$. Lag plots exhibiting no correlation suggest that the residuals do follow a random normal distribution. Lag plots with correlated (positive or negative) residuals indicate some degree of auto-correlation. The $DW$ statistic is given as,
\begin{center}
        \begin{equation}
        \textit{DW} = \frac{\sum_{i=2}^{N}[R_i - R_{i-1}]^2}{\sum_{i=2}^{N}[R_i]^2}
        ,\end{equation}
\end{center}

where $R_i$ is the residuals in the original order and $R_{i-1}$ is the $k = 1$ lagged residual. The range of $DW$ is 0 to 4 with the two extrema corresponding to anti-correlation and auto-correlation, respectively, and a $DW$ = 2 indicates no auto-correlation in the investigated residuals.

\begin{center}
        \begin{figure*}
                \centering
                \includegraphics[width=500px]{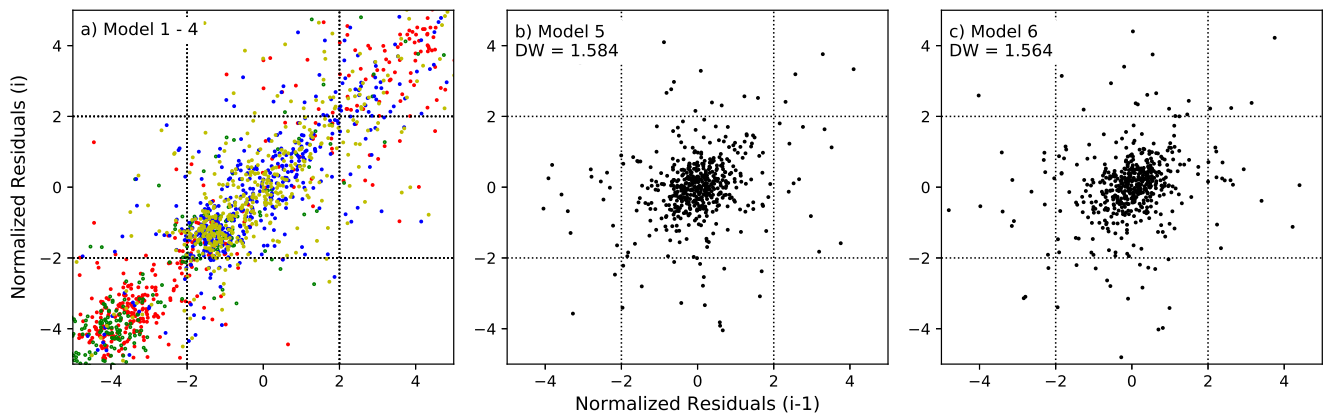}
                \caption{Lag plots corresponding the six models considered in this study. Lag plots a) for Models 1 - 4 (red, green, blue, and yellow, in respective order), b) for Model 5, and c) for Model 6. Plots b) and c) show the Durbin-Watson ($DW$) statistic. We refer the reader to the text for more details.}
                \label{figure:lag_plots}
        \end{figure*}
\end{center}

With the increased model complexity in our analysis, the $DW$ statistic converges to 2 ($DW$ $\approx$ 2) and the auto-correlation seems to diminish (Fig. \ref{figure:lag_plots}). The lag plots of Models 1 - 4 show clear signs of deviation from normal distribution around the origin. The $DW$ statistics of Models 1 - 4 are, 0.055, 0.057, 0.326, and 0.340, in respective order, all of which are far from $DW = 2$. Adding a $\beta$ parameter has no significant effect on either the $DW$ statistic or the auto-correlation of the residuals, when comparing Models 1 and 2 as well as Models 3 and 4.

The models with two LiTEs have almost the same $DW$ statistics, while Model 5 has a $DW$ statistic slightly closer to 2. However, the difference is so small that any conclusion from $DW$ statistics cannot be achieved. The auto-correlation of the residuals in lag plots for Models 5 and 6 looks almost the same, and these are both close to being normally distributed, as can be seen in the figure. With $DW$ values deviating from 2, Models 5 and 6 can be interpreted as follows: either $i)$ an additional trend of likely astrophysical origin is present if the timing errors distribute normally, or $ii)$ the deviation from $DW$ = 2 is due to measurement errors and/or the presence of additional astrophysical signals in the data if timing errors are not normally distributed.

The ETV dataset of HW Vir consists of observations performed in different atmospheric conditions with various instruments having different characteristics. Mid-eclipse times are also measured following different methods from data reduced making use of different packages. Moreover, variable atmospheric conditions can introduce time-correlated, so-called red noise, even on the observations from the same instrument. All of these factors, and perhaps others, may be responsible for timing errors not being normally distributed (for a detailed discussion on potential error sources, see \cite{2016arXiv160703680V}). Even if the observational errors of ETV data are not normally distributed, we cannot discard the possibility of assuming wrong ETV models and/or additional astrophysical signals over assumed models on the ETV.

\subsection{Model comparison: $\chi^2$ test}\label{modelselectionchi2test}

We compared the $p$-value, $p(\chi^2,\nu)$, of each model calculated from an $\chi^2$ distribution for a given number of degrees of freedom, with a critical probability indicated by the significance level ($\alpha$) to test the null-hypothesis stating that the model under investigation fits the data well ($p < \alpha$). In this case, the probability of finding the observed $\chi^2$ is too low at the given significance level and cannot be explained by a random process. When $\alpha < p \leq 0.5$, the discrepancies between the model and the data are random; there is no evidence to reject the null-hypothesis. Finally, if $0.5 < p < 1$, the standard error used in the determination of $\chi^2$ is overestimated, resulting in an unrealistically small value of $\chi^2$ \citep{Hughes2010}.

At the heart of a statistical significance test is the assumption that data are independent and normally (random) distributed. Assuming the null-hypothesis, the expectation value of $\chi^2$ is close to the mean of the ($\chi^2, \nu$) distribution: $\chi^2 \approx \nu$ with probability $p(\chi^2, \nu) \approx 0.5$. In the case of considerably large $\chi^2$ values, the model function that we assumed to fit the data well in the null-hypothesis is unlikely to explain the observations, which can be quantitatively decided by comparing the probability of the $\chi^2$ to a critical probability, $\alpha$. If the $\chi^2$ is considerably less than the mean value of the $\chi^2$ distribution, the probability of the $\chi^2$ becomes close to unity, which eventually means the standard errors of the data are overestimated; in other words, we are fitting the noise of the data.

The critical probability, alpha, sets the significance level (1-alpha) and is somewhat  arbitrary depending on the nature of the underlying problem. Often $\alpha$ is chosen to be  0.001 (0.1\%), 0.01 (1\%), or 0.05 (5\%). The significance levels of rejection and non-rejection are 99.9\%, 99\%, and 95\%, respectively. The $p$-value itself provides a measure of the strength of the evidence against the null-hypothesis. As a rough guideline, we adopt the following criteria: if $p(\chi^2) < 0.01,$ we judge very strong evidence against $H_0$; if  $0.01 < p < 0.05,$ we judge strong evidence against $H_0$; if $0.05 < p < 0.10,$ we judge some weak evidence against $H_0$; and if $p > 0.10,$ we judge little or no evidence against $H_0$.

Table \ref{table:chi2test_results} shows the calculated $\chi^2$ for each model that we considered in this study. To calculate the probabilities $p(\chi^2, \nu)$ for each model, we made use of our own {\sc PYTHON} code. For Models 1 - 4, the corresponding $p$-values are smaller than $10^{-200}$, thus with a high level of significance, and under the mentioned assumptions, we can reject the null-hypothesis for these models. 

For Models 5 and 6, we find the $p$-value to be 0.285 and 0.155, respectively. Considering the significance level ($\alpha$ = 0.1), we see no evidence against Models 5 and 6, since $p_{model} > \alpha$. In a relative sense, we  therefore interpret this result as a non-rejection of the null-hypothesis for Models 5 and 6.

\begin{center}
        \begin{table}
                \centering
                \begin{tabular}{lcccc}
                        &\bm{$\nu$}&\bm{$\chi^2$}&\bm{$\chi_{\nu}^2$}&\bm{$p$}\textbf{-value}\\ [3pt]
                        \hline \\ [-5pt]
                        Model 1&$507$&$45263.8$&$89.28$&$< 10^{-300}$\\
                        Model 2&$506$&$40236.1$&$79.52$&$< 10^{-300}$\\
                        Model 3&$502$&$2801.1$&$5.58$&$< 10^{-300}$\\
                        Model 4&$501$&$2633.0$&$5.26$&$2.2 \times 10^{-285}$\\
                        Model 5&$497$&$514.5$&$1.04$&$0.285$\\
                        Model 6&$496$&$528.0$&$1.07$&$0.155$\\
                        \hline \\ [-5pt]
                \end{tabular}
                \caption{Results of the $\chi^2$-test ($\alpha$ = 0.1; see text for more details).}
                \label{table:chi2test_results}
        \end{table}
\end{center}

\subsection{Model selection: $F$-Test}\label{ftest}
In the previous section, we determined the most likely model by comparing the minimum $\chi^2$ of each model with the $\chi^2$ probability density distribution for $\nu$ degrees of freedom under the assumption of the null-hypothesis. We find an $\chi^2$ value for the Model 5, indicating the greatest probability amongst other candidates, and hence deem it to be very significant not to reject the null-hypothesis based on the $p$-value. While evaluating a low $\chi^2$ is a necessary requirement for assessing the goodness of fit, this method also has limitations favoring models overfitting the data. The more the adjustable parameters are included, the better the overall goodness of fit will be (lower variance). Therefore, one drawback of the $\chi^2$ test is that it does not have a mechanism to distinguish models with increasing numbers of adjustable model parameters (for the same number of data points).

In this section, we describe the application of the $F$-test to our problem and report the results. This test aims to quantitatively select one model in favor of another model. Intuitively, a model (2) with more adjustable parameters should describe a given dataset at least as well as a model (1) with fewer adjustable parameters. Hence, model 2 will provide lower residual errors. The question is then: how significant is the improvement of model (2) over model (1)? The test statistic quantifying this problem is given as

\begin{center}
        \begin{equation} \label{eq:f_value}
        \textit{F} = \frac{(\chi^2_1 - \chi^2_2) / (\nu_2 - \nu_1)}{\chi^2_2 / \nu_2}
        ,\end{equation}
\end{center}

where $\nu_i$ are the corresponding degrees of freedom, and $\chi^2_i$ are the $\chi^2$ statistics for any two models that are being compared.

With the null-hypothesis stating that the model (2) does not describe the data significantly better than model (1), the F-statistic follows the F probability density distribution. To assess the rejection probability, we again used the calculated $p$-value for the observed F-statistic as outlined in the previous section.

By using our own {\sc PYTHON} code, we calculated F-statistics as given in Eq. \ref{eq:f_value} by adopting $\chi^2$ and $\nu$ values from the best-fit results from Sect. \ref{etvmodeling} (see Table \ref{table:chi2test_results}), for each model pair ordered by the corresponding $\nu$. We then calculated the probability of F-statistics in an $F$-distribution defined by the two degrees of freedom, $P(F;v_2-v_1,v_2)$.  Similar to the $\chi^2$ test, we chose the critical probability, $\alpha$, as 0.1 (10\%), and the condition of rejecting the null-hypothesis is in the cases of $p < \alpha$, while if $p > \alpha$, there is no reason to state that the simpler model does not fit the data statistically well compared to the more complex one. The model pairs and their corresponding $F$-statistics and $p$-values can be seen in Table \ref{table:ftest_results}.

Other than for Models 5 and 6, $p$-values of all of the remaining $F$-tests were closer to zero than being comparable to $\alpha$ of 0.1, thus the null-hypotheses for all of these cases were rejected. $F$-value being negative for the test between Models 5 and 6 means that not only did the additional parameter not improve the fit, but also the more complex model has poorer statistics than the simpler one. Therefore, the $p$-value is unity and there is no reason for rejecting the null-hypothesis.

In all of the model selection and comparison techniques performed in this study, Models 5 and 6 are significantly better compared to the other simpler models. The statistical significance of Model 5 over Model 6 could be misleading. Such orbital architecture is also expected to be unstable even within a few orbital revolutions. Nevertheless, we checked the stability of Model 5 in Sect. \ref{stability}. From a physical point of view, we expect that Model 5 is not feasible. On the other hand, the additional objects of Model 6 have much smaller minimum masses with a relatively larger separation between the two. The observational signature of these two bodies may expected to be undetectable, if their orbital inclinations are relatively close to $90^\circ$, thus their true masses are close to the minimum masses. The eccentricities are larger compared to Model 5 and should be subject to the dynamical analysis. However, we do not see a direct reason to reject Model 6. Therefore, we adopt Model 6 as a possible explanation to the ETV of HW Vir.

\begin{center}
        \begin{table}
                \centering
                \begin{tabular}{lcc}
                        \bm{$F$}\textbf{-test between}&\bm{$F$}\textbf{-value}&\bm{$p$}\textbf{-value} \\ [3pt]
                        \hline \\ [-5pt]
                        Model 1 - Model 2&$63.2$&$1.2\times10^{-14}$\\
                        Model 2 - Model 3&$1677.3$&$7.8\times10^{-289}$\\
                        Model 3 - Model 4&$32.0$&$2.6\times10^{-8}$\\
                        Model 4 - Model 5&$511.7$&$1.2\times10^{-174}$\\
                        Model 5 - Model 6&$-12.8$&$1$\\
                        \hline \\ [-5pt]
                \end{tabular}
                \caption{Results of the $F$-test (see text for more details).}
                \label{table:ftest_results}
        \end{table}
\end{center}

\section{Error analysis}\label{erroranalysis}

Preliminary uncertainties for each parameter of Model 6 have already been assigned as shown in Table \ref{table:OCresultsHWVir}. However, these are formal errors because they have been derived from the covariance matrix as a result of the {\sc LMFIT} least-squares minimization process. In general, formal uncertainties are not reliable, since parameter correlations are not taken into account. 

In order to compute realistic parameter uncertainties for Model 6, we made use of two different and independent algorithmic methods as part of a comparative analysis study. The first method explores the projected $\chi^2$ surface for any two parameters. We believe this method is somewhat overlooked in the astronomical data analysis work, and we provide an in-depth description of the underlying machinery. The second method is based on the bootstrap Monte Carlo resampling method, which is widely practiced in data analysis in astronomy.

\subsection{$\chi^2$ surface search}\label{chi2surfacesearch}

Parameter uncertainties can be determined by exploring the projected $\chi^2$ surface around $\chi^2_{min}$. In the following, we chose to adopt the technique outlined in \citet[p.74]{Hughes2010}. The 68.3\% confidence level for a given parameter is calculated numerically by successive iterations until the condition $\chi^2_{min}$ + 1 is met. This method is then applied to two-parameter projections of $\chi^2$ in order to account for the parameter correlations that are otherwise neglected when quoting formal uncertainties. For a systematic change of any two parameters around their best-fit values we monitored the $\chi^2$ value while letting the remaining parameters to vary freely. The upper and lower standard deviations were then determined from $\chi^2_{min}$ + 1. Any parameter correlation is propagated and accounted for in the calculated confidence intervals. For complicated correlations, the $\chi^2$ topology leads to asymmetric error bars around the best-fit values \citep{Bevington2003}.

There are 13 parameters in the analytic expression of Model 6. This corresponds to a total of 78 parameter pairs. We selected each pair within these sets and fixed them to 75 different values around the best-fit model considering appropriate ranges. Thus, the resolution of the $\chi^2$ surface was $75~\times~75,$ and we calculated a total of 438750 fits. For each fit, we calculated and stored the resulting $\chi^2$ corresponding to a given parameter pair. To map out the $\chi^2$ topology in some more detail, we chose to adopt an exponential parameter gradient around the best-fit model. This approach provides a higher resolution of the $\chi^2$ landscape for values close to the best-fit model and a lower resolution far from it. The reason for this choice is to reduce the computation time and assign weights to the topology change in the vicinity of $\chi^2_{min}$. For each grid point, we stored the resulting $\chi^2$ value.

We converted the resulting $\chi^2$ surfaces to a grid of probability density as $P(x,y) \propto exp(-\chi^2/2)$, where the latter quantity is the maximum likelihood and $x$ and $y$ represent the two parameters. We normalized the probabilities to impose the constraint $\int P(x,y)dxdy = 1$. We created 12 marginalized probability density distributions for each parameter by summing the probability values that correspond to the same parameter value on the grid of each pair. We then summed these 12 probability density distributions to create a final normalized probability distribution for 13 parameters. Finally, we calculated the 68.3\% confidence intervals as well as median values from the final probability distributions.

In addition to the model parameters, we calculated derived parameters of the mass function ($f(m)$), minimum mass ($M\sin\!i$), and projected semi-major axis ($a\sin\!i$). We calculated their confidence intervals by following the functional approach of error propagation for multi-variable functions \citep[see][]{Hughes2010}. The procedure is relatively straightforward, but there is merit in some detailed explanation for the purpose of clarity. The derived parameter is defined as  X = $f(\alpha, \beta, ...)$, where ($\alpha, \beta, ...$) are the median values of parameters determined from the marginalization process. To obtain a conservative estimate we then added the upper limit in uncertainty to the median of a given parameter and calculated the difference $f(\alpha + \sigma_{\alpha}, \beta, ...) - f(\alpha, \beta, ...)$. This procedure and calculation of differences were also determined for the other parameters $f(\alpha, \beta + \sigma_{\beta}, ...) - f(\alpha, \beta, ...)$. The final uncertainty for X is then obtained from the quartiles as

\begin{multline*}
        \sigma_{X}^2 = [f(\alpha \pm \sigma_{\alpha}, \beta, ...) - f(\alpha, \beta, ...)]^2  \\
        + [f(\alpha, \beta \pm \sigma_{\beta}, ...) - f(\alpha, \beta, ...)]^2 + ...
\end{multline*}

Since the uncertainties calculated by the $\chi^2$ surface search are not necessarily normally distributed and their distribution can be asymmetric, we needed to calculate the uncertainties both for the positive and negative sides separately by using the formula above. According to \citet[ p.~41]{Hughes2010}, this procedure of error propagation assumes the variables that are needed to calculate the derived parameters to be uncorrelated and independent. To determine the derived parameters, we adopted the uncertainties derived from 68.3\% confidence intervals of $\chi^2$ surface. Therefore, the propagated uncertainties are not formal and the effects of correlations on the error surface were already taken into account. The normalized probability distributions for each parameter are shown in Fig. \ref{figure:chi2surface_probabilities} in the appendix, and the values that correspond to the 68.3\% confidence levels are shown in Table \ref{table:errorsearchresults}. $\chi^2$ surface plots are shown in Appendix \ref{appendixchi2surface}.

\subsection{Bootstrap method}\label{bootstrap}

Another method to determine parameter uncertainties is known as the re-sampling bootstrap method \citep{article}. Again, we aim to determine the parameter uncertainties in Model 6. 

The boostrap method is also referred to as the 'quick and dirty Monte Carlo method' in \cite{1992nrca.book.....P}, who found wide-spread acceptance in frequentist astrostatistics to make inferences of the parent distribution for each parameter. The derived uncertainties can then be compared to the uncertainties obtained from the marginalization process as described in the previous section. The underlying algorithm is based on random sampling from the original dataset with a replacement. A given simulated (bootstrapped) dataset has the same number of data points as the original dataset. The replacement condition introduces the possibility that data points can be duplicated in a given bootstrap set. Since the dimension of the dataset is equal, this implies that some data is missing among simulated datasets.

There is, however, one question remains to be addressed concerning the number of random samples. To answer this question, we performed a systematic numerical experiment where we calculated parameter distributions for a range of the number of random samples. We considered $1 \times 10^4$, $2.5 \times 10^4$, $1 \times 10^5,$ and $2 \times 10^5$ random samples and plotted the resulting distributions. We found no difference between $1 \times 10^5$ and $2 \times 10^5$ bootstrap samples. Clearly, choosing $1 \times 10^4$ is too small a number to warrant a reliable large-number statistic of the final result.

As described in the previous section, we again determined the 15.9\%, 50\%, and 84.2\% quartiles from each parameter distribution in order to obtain the median and 68.2\% confidence interval. In Fig. \ref{figure:bootstrap_histogram} in the appendix, we give the histograms obtained from the bootstrap method. Table \ref{table:errorsearchresults} summarizes the uncertainties for Model 6 parameters. We would like to emphasize the relative strength of the bootstrap method, which is proven to work on any parent distribution, even without knowing the nature of it at all. Hence, we recommend taking the usage of uncertainties from bootstrap into account in any related study in order to be cautious in further interpretation rather than stating only the formal errors.

\begin{center}
        \begin{table*}
                \centering
                \begin{tabular}{cccccc}
                        &\textbf{Parameter}&\textbf{Unit}&\textbf{$\chi^2$ Surface Search} &\textbf{Bootstrap}&\textbf{Agreement Level ($\sigma$)}\\
                        \hline\\[-6pt]
                        \multirow{2}{*}{Lin}&$T_{0}$&$BJD_{TDB}$        &$2,450,280.2857\ _{-1.2 \times 10^{-4}}^{+1.5 \times 10^{-4}}$   &$2,450,280.2857\ _{-1.3 \times 10^{-4}}^{+2.2 \times 10^{-4}}$&$0.068$\\[4pt]
                        &$P_{0}$&$d$                                            &$0.116719556\ _{-7.4 \times 10^{-9}}^{+7.3 \times 10^{-9}}$   &$0.116719556\ _{-4.45 \times 10^{-9}}^{+4.73 \times 10^{-9}}$&$0.001$\\[4pt]
                        \hline\\[-6pt]
                        Quad&$\beta$&$d/cycle^{2}$                      &$-1.42 \times 10^{-12}$$\ _{-1.9 \times 10^{-13}}^{+1.6 \times 10^{-13}}$       &$-1.46 \times 10^{-12}$$\ _{-2.02 \times 10^{-13}}^{+1.35 \times 10^{-13}}$&$0.171$\\[4pt]
                        \hline\\[-6pt]
                        \multirow{5}{*}{$LiTE_1$}&$T_{0,1}$&$BJD_{TDB}$ &$2,444,243\ _{-776}^{+407}$                         &$2,444,230\ _{-798}^{+266}$                    &$0.012$\\[4pt]
                        &$P_1$&$d$                                                                      &$11494\ _{-483}^{+691}$                         &$11310\ _{-356}^{+801}$                        &$0.174$\\[4pt]
                        &$e_1$&-                                                                        &$0.450\ _{-0.011}^{+0.011}$             &$0.448\ _{-0.020}^{+0.021}$            &$0.091$\\[4pt]
                        &$A_1$&$s$                                                              &$155\ _{-14}^{+22}$                           &$154\ _{-13}^{+20}$                            &$0.064$\\[4pt]
                        &$\omega_1$&$^{\circ}$                                                  &$359\ _{-2}^{+2}$                                     &$0\ _{-5}^{+5}$                                        &$0.070$\\[4pt]
                        &$f(m)_{1}$&$M_{jup}$                                           &$0.045\ _{-0.018}^{+0.022}$             &$0.045\ _{-0.011}^{+0.021}$            &$0.012$\\[4pt]
                        &$M_{1} \sin\!i$&$M_{jup}$                                              &$25.1\ _{-2.5}^{+3.7}$                         &$25.0\ _{-2.2}^{+3.5}$                         &$0.013$\\[4pt]
                        &$a_{1} \sin\!i$&$au$                                                   &$7.9\ _{-1.1}^{+1.5}$                                 &$7.8\ _{-1.0}^{+1.4}$                          &$0.041$\\[4pt]\hline
                        \\[-6pt]
                        \multirow{5}{*}{$LiTE_2$}&$T_{0,2}$&$BJD_{TDB}$ &$2,452,958\ _{-80}^{+78}$                   &$2,452,939\ _{-107}^{+101}$                    &$0.137$\\[4pt]
                        &$P_2$&$d$                                                                      &$4963\ _{-31}^{+32}$                           &$4949\ _{-68}^{+74}$                           &$0.167$\\[4pt]
                        &$e_2$&-                                                                        &$0.271\ _{-0.022}^{+0.022}$             &$0.279\ _{-0.025}^{+0.026}$            &$0.212$\\[4pt]
                        &$A_2$&$s$                                                              &$54\ _{-1}^{+1}$                                     &$54\ _{-1}^{+2}$                                       &$0.066$\\[4pt]
                        &$\omega_2$&$^{\circ}$                                                  &$13\ _{-6}^{+6}$                                     &$12\ _{-8}^{+8}$                                       &$0.139$\\[4pt]
                        &$f(m)_{2}$&$M_{jup}$                                           &$0.00798\ _{-0.00048}^{+0.00051}$         &$0.00802\ _{-0.00070}^{+0.00103}$      &$0.029$\\[4pt]
                        &$M_{2} \sin\!i$&$M_{jup}$                                              &$13.9\ _{-0.3}^{+0.3}$                         &$13.9\ _{-0.45}^{+0.60}$                       &$0.029$\\[4pt]
                        &$a_{2} \sin\!i$&$au$                                                   &$4.57\ _{-0.16}^{+0.16}$                       &$4.56\ _{-0.22}^{+0.27}$                       &$0.026$\\[2pt]
                        \hline\\[-8pt]
                \end{tabular}
                \caption{Results of best-fit parameters and associated uncertainties for Model 6 as determined from the $\chi^2$ surface search and bootstrap method. Further details are given in the text. The agreement level is calculated as $ABS(X-Y)$ / $\sqrt{dX^2 + dY^2}$. Results of an $\chi^2$ surface search and bootstrap search with their agreements in units of $\sigma$.}\label{table:errorsearchresults}
        \end{table*}
\end{center}

\section{Stability of Model 6}\label{stability}

The orbital parameters of Model 6 derived from the ETV analysis correspond to two planetary-mass companions in eccentric orbits that overlap.  The additional companions thus undergo strong mutual gravitational interactions, and system stability is only possible if a resonant mechanism is present to prevent close encounters. Therefore, a dynamical stability search of the orbits derived from Model 6 is needed.

We performed a stability search for the results from the LM fit, a bootstrap search, and an $\chi^2$ surface search. We used frequency map analysis \citep[FMA;][]{1990Icar...88..266L,1993PhyD...67..257L}, similarly to the studies of \cite{2010A&A...511A..21C} and \cite{2010A&A...519A..10C}. We calculated the normalized stability index $D$ as
\begin{center}
        \begin{equation} \label{eq:stability_logD}
        \textit{$D$} = \frac{\lvert n_1 - n_2\rvert}{n_1}
        ,\end{equation}
\end{center}
where $n_1$ is the frequency of the mean motion calculated from the first half of each integration, and $n_2$ is the same for the second half. In the case of regular motion, the normalized stability index should be $D < 10^{-6}$ \citep{2005A&A...440..751C}.

For the integrations, we used the MERCURY6 code \citep{1999MNRAS.304..793C} with the RADAU algorithm. We set the timestep as 100 d and total integration time as 3 $\times 10^6$ d. We calculated $n_1$ and $n_2$ frequencies using a TRIP code \citep{Gastineau:2011:TCA:1940475.1940518}. We integrated the system by varying the semi-major axis and eccentricity of the outer companion while keeping the orbital parameters of the inner companion fixed since the uncertainties on the inner companion are smaller compared to those in outer orbit. 
The orbital period of the inner circumbinary orbit is more than $4 \times 10^4$ longer than the orbital period of the central binary. In this case, interactions between an individual star and a circumbinary object can be neglected. Therefore, we assumed that the central binary as a single object with a mass of the total binary mass. 
We also assumed that the orbits of the two circumbinary objects are coplanar and inclinations are $90^\circ$. Therefore, we fixed the masses and the other orbital parameters to the values in Table \ref{table:errorsearchresults}. We set the step size for $a$ as 0.04 $au$ and 0.005 for $e$.

\begin{center}
        \begin{figure}
                \centering
                \includegraphics[width=240px]{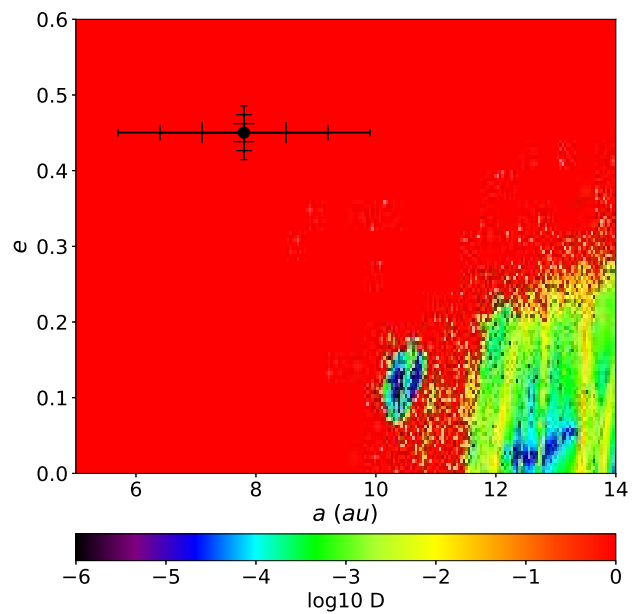}
                \caption{Stability analysis of Model 6 in Sect. \ref{etvmodeling} (Table \ref{table:OCresultsHWVir}). The color scale represents the normalized stability index in log scale. The black circle represents the best-fit values of the outer orbit. The errorbars represent 1, 2, and 3 $\sigma$ uncertainties on the best-fit values (see text for more details).}
                \label{figure:fma_lmfit}
        \end{figure}
\end{center}

\begin{center}
        \begin{figure}
                \centering
                \includegraphics[width=240px]{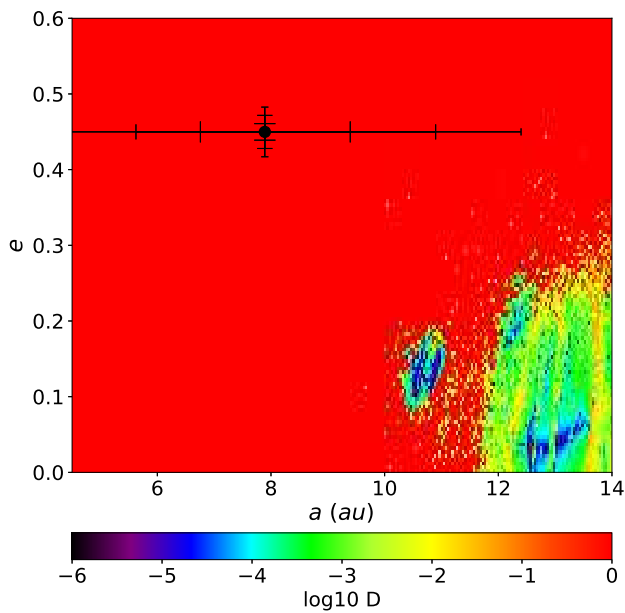}
                \caption{Stability analysis of Model 6 from $\chi^2$ surface search in Sect. \ref{chi2surfacesearch} of Model 6. See Fig. \ref{figure:fma_lmfit} for the description of the figure (see text for more details).}
                \label{figure:fma_chi2surface}
        \end{figure}
\end{center}

\begin{center}
        \begin{figure}
                \centering
                \includegraphics[width=240px]{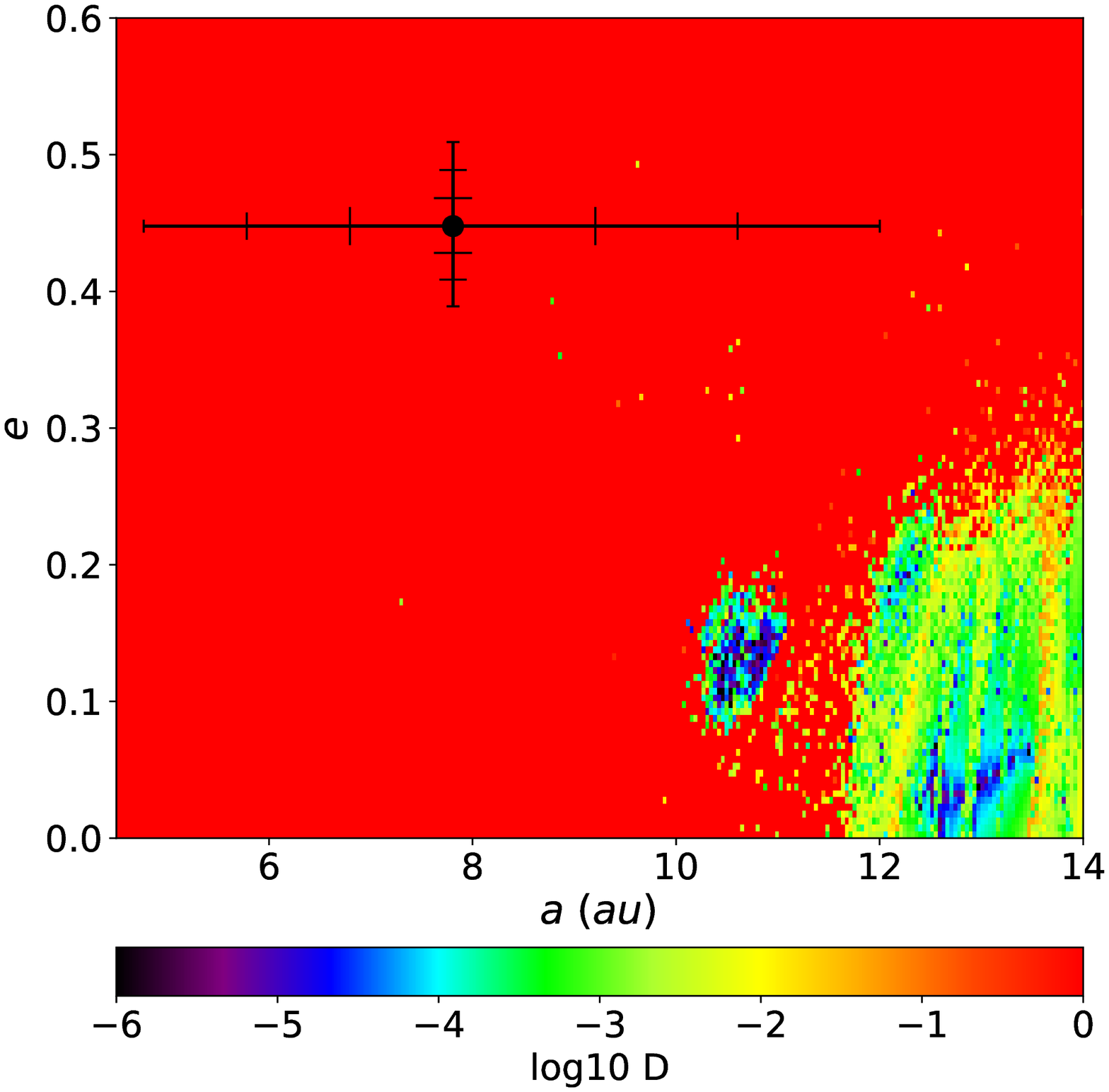}
                \caption{Stability analysis of Model 6 from bootstrap method in Sect. \ref{bootstrap} of Model 6. See Fig. \ref{figure:fma_lmfit} for the description of the figure (see text for more details).}
                \label{figure:fma_bootstrap}
        \end{figure}
\end{center}

Almost all of the orbits within the searched $a$ - $e$ range are unstable other than a small stable region ($D < 10^{-6}$) around $a \sim 10.5 \ au$, $e \sim 0.12$ and another one around $a > 12.2 \ au$, $e < 0.1$ (Figs. \ref{figure:fma_lmfit}, \ref{figure:fma_chi2surface}, and \ref{figure:fma_bootstrap}). None of the orbital configurations of Model 6 are in stable regions within a 3$\sigma$ uncertainty range. In the presence of the inner companion with the parameters derived from Model 6, the outer companion should have stable orbits for $a$ > 10 $au$ and $e$ < 0.2 (black regions in Fig. \ref{figure:fma_lmfit}, \ref{figure:fma_chi2surface} and \ref{figure:fma_bootstrap}). We tried to fit the ETV by using the parameters that correspond to the stable regions mentioned above. However, none of these trials corresponded to satisfactory fits. Therefore, we conclude that Model 6 is not plausible in terms of explaining the ETV of the HW Vir system.

We also checked if the circumbinary stellar mass objects of Model 5 have any stable orbits. By assuming the central binary as a single object and keeping the two orbits coplanar, we simulated the best-fit orbital configuration of Model 5. As expected, such massive objects are interacting with each other very strongly, and the outer object is being ejected in less than one orbital revolution. Therefore, Model 5 can be ruled out as well.

\section{Conclusions}\label{conclusion}

From the simultaneous light and velocity curve analysis in Sect. \ref{lightcurveanalysis}, mass of the primary is found to be 0.413 $M_{\sun}$ and 0.307 $M_{\sun}$ for the WS99+H96 and ED08 datasets, respectively. Both of the masses are below the canonical helium ignition mass of $\sim$0.47 $M_{\sun}$. However, similar values below the canonical mass have been given in the literature previously. \cite{1986IAUS..118..305M} found an even smaller mass for the primary companion as 0.25 $M_{\sun}$. A recent study of \cite{2018MNRAS.481.2721B} used the method of \cite{2010ApJ...717L.108K}, which makes use of LiTE, and derived a mass of 0.26 $M_{\sun}$ for the primary of HW Vir. On the other hand, \cite{2008ASPC..392..187E} found a larger value for the mass of the primary as 0.53 $M_{\sun}$. This wide range of the masses of the sdB component in the HW Vir system is still an open question requiring further investigation, which is beyond the scope of this work. HW Vir is a challenging case for light curve analysis due to its compact orbital configuration and high contrast in the luminosities of the individual components. We do not intend to speculate about whether there is a core helium ignition process in the primary, since our intention is to investigate eclipse timing variation of the binary.

Due to its very low luminosity compared to the sdB primary, little is known about the secondary star. We found 0.128 $M_{\sun}$ and 0.110 $M_{\sun}$ in our analysis based on the WS99+H96 and ED08 datasets, respectively, for the mass of the secondary. Literature values for the secondary mass change between 0.1 $M_{\sun}$ from \cite{2018MNRAS.481.2721B} and 0.15 $M_{\sun}$ from \cite{2008ASPC..392..187E}. Almost all these masses are within the range of a main-sequence M star. However, the temperature values that we found are somewhat higher for an M-type main-sequence star, probably due to the reflection effect. For both WS99+H96 and ED08 datasets, we found surface temperatures of $\sim$3900 K for the secondary. The properties of the secondary companion in HW Vir, as well as many other secondaries in sdB+dM binaries, will be tentative until precise spectrographic observations can be obtained to distinguish it from those of the hot primary. The sizes of the primary and the secondary stars are found to be almost the same for both of the radial velocity datasets, which is consistent with the partial primary and secondary eclipses, in spite of the tight orbital configuration ($a \approx 0.8 \ R_{\sun}$) observed under rather high inclination ($i \approx 81^{o}$).

Our ETV analysis for the seasonally binned data gave almost the same results as the original unbinned data. Since the binning size is considerably shorter than the cyclic trends on ETVs, we did not encounter any loss of information; for example, LiTE amplitude, period etc. Moreover, taking weighted averages of the data minimized the effects of the outliers, which can impact the results of the least-squares minimization methods \citep{Mandel1964}. From our results, we can conclude that using seasonally binned data is permitted for the case of HW Vir.

The $\beta$ coefficient of Model 6 in the ETV analysis corresponds to a constant period change with a linear rate of $-9.16 \times 10^{-9} d \ yr^{-1}$. The angular momentum change can be derived \citep[e.g.,][]{2006MNRAS.365..287B} as $dJ/dt = -3.25 \times 10^{35} erg$. The 3$\sigma$ confidence interval of the angular momentum change should be within the range of $-4.60 \times 10^{35} erg$ and $-2.36 \times 10^{35} erg$. The possible mechanisms of angular momentum loss may be gravitational radiation or magnetic stellar wind breaking. By using the absolute parameters derived in Sect. \ref{lightcurveanalysis} and the formulation from  \cite{1967AcA....17..287P}, the gravitational radiation of the binary should be $-6.57 \times 10^{32} erg$. The angular momentum loss calculated for the $\beta$ coefficient is three orders of magnitude larger as it is caused by gravitational radiation. On the other hand, magnetic stellar wind breaking due to the secondary star of the HW Vir system will be within the range of $-5.48 \times 10^{36}$ and $-4.16 \times 10^{33}$ \citep{1983ApJ...275..713R}, which makes it possible to explain the rate of constant period change in ETV.

The circumbinary objects from Model 6, in a wide range of $a$ and $e$ parameters are found to be unstable, unless the outer object has $a$ > 10 $au$, for which no ETV solution is found. The stable configuration that \cite{2012A&A...543A.138B} found was shown to diverge from the more recent timing data in \cite{2018MNRAS.481.2721B}. Therefore, circumbinary objects alone cannot explain the ETV periodicities. One should also keep in mind that the analytical expression of LiTE \citep{1959AJ.....64..149I} does not consider the mutual interactions in a system that consists of more than two bodies. A better solution would be to analyze the ETV by simultaneously solving n-body interactions between the bodies in the case of multiple circumbinary objects.

Another explanation for the apparent ETV trend might be the effect of the magnetic activity modulation of the cooler companion. Our initial sinusoidal fit to the ETV resulted in a period of 77 yr. By using the formulation of \cite{1992ApJ...385..621A}, we calculate the magnetic field as $B \approx 39 \ kG$. The luminosity variation of the secondary with $L_2 = 0.006\ L_{\odot}$ should be $\Delta L = 0.011\ L_{\odot}$. Other than the amount of change being larger than the luminosity level of the secondary, such a change was not reported by any observer. Therefore, magnetic activity of the secondary can be ruled out at the level of period change calculated in this study. However, the possibility of post-common envelope evolution and/or a close orbital configuration of the binary affecting the magnetic activity of the cooler companion cannot be ruled out. We would like to encourage interested researchers to investigate the cooler companion and its effect on ETVs in HW Vir-like binaries.

The fit statistics and parameter uncertainties heavily depend on timing uncertainties. \cite{2014CoSka..43..382M} questioned the reliability of the uncertainties from the widely used Kwee-van Woerden method \citep{1956BAN....12..327K} in determining the eclipse timings. From their comparison with the least-squares method, they concluded that the uncertainties from the Kwee-van Woerden method are systematically underestimated. Analyses based on underestimated uncertainties may result in the underestimation or overestimation of the parameter uncertainties depending on the model, if they are not carefully examined. Therefore, we suggest carefully investigating error estimates on the mid-eclipse timings.

As \cite{2018MNRAS.481.2721B} showed, the primary star in the HW Vir system displays pulsations that are not detectable from the ground. Since many proposed circumbinary planets are in systems with components expected to show pulsations with different timescales and amplitudes, the effects of pulsations on the measurement of the eclipse timings should also be investigated, even when they only contribute to the noise budget, especially in ground-based observations. 

\begin{acknowledgements}
        We thank TÜBİTAK National Observatory for a partial support in using T100 telescope with the project number 17BT100-1208 and 17BT100-378. We also thank Ankara University Kreiken Observatory (AUKR) for the observation time and all the student observers who helped in the observations of the target. E.M.E. acknowledges support from TÜBİTAK (2214-A, No. 1059B141800521). O.B. thanks The Scientific and Technological Council of Turkey (TÜBİTAK) for their support through the research grand 118F042. T.C.H acknowledges staff at SOAO and LOAO observatories for assistance with the observations and fruitful discussions. T.C.H acknowledges financial support from the National Research Foundation (NRF; No. 2019R1I1A1A01059609). A.C. acknowledges support by CFisUC projects (UIDB/04564/2020 and UIDP/04564/2020), GRAVITY (PTDC/FIS-AST/7002/2020), ENGAGE SKA (POCI-01-0145-FEDER-022217), and PHOBOS (POCI-01-0145-FEDER-029932), funded by COMPETE 2020 and FCT, Portugal.
\end{acknowledgements}

\nocite{*}

\bibliography{hwvir_ref}
\bibliographystyle{aa}


\begin{appendix} 

\section{Mid-Eclipse Timings}\label{table:mideclipse_times}
\textcolor{white}{space}
\tablefirsthead{\multicolumn{1}{c}{\textbf{Mid-Eclipse Times}} & \multicolumn{1}{c}{\textbf{Error}} & \multicolumn{1}{c}{\textbf{Type}} & \multicolumn{1}{c}{\textbf{Filter}} & \multicolumn{1}{c}{\textbf{Tel.}}\\
\bm{$(BJD_{TDB})$}&\multicolumn{1}{c}{\textbf{(d)}}&\textbf{p/s}&&\\
\hline\\
}
\topcaption{List of new primary (p) and secondary (s) eclipse timings as presented in this work (see text in Sect. \ref{timingdatamodeling} for details).}
\begin{supertabular}{clccc}
        \label{table:minimaHWVir}
        2456708.494970& $\pm$0.000024   &p      &R      &TUG\\
        2456708.553380& $\pm$0.000072   &s      &R      &TUG\\
        2456708.611720& $\pm$0.00002    &p      &R      &TUG\\
        2456813.367760& $\pm$0.000095   &s      &R      &TUG\\
        2457075.461120& $\pm$0.000063   &p      &R      &AUKR\\
        2457095.478600& $\pm$0.00013    &s      &R      &AUKR\\
        2457095.536860& $\pm$0.000081   &p      &R      &AUKR\\
        2457103.123850& $\pm$0.000092   &p      &R      &SOAO\\
        2457103.182120& $\pm$0.00014    &s      &R      &SOAO\\
        2457103.240600& $\pm$0.000054   &p      &R      &SOAO\\
        2457103.298927& $\pm$0.00036    &s      &R      &SOAO\\
        2457104.115700& $\pm$0.000042   &s      &R      &SOAO\\
        2457104.174010& $\pm$0.000049   &p      &R      &SOAO\\
        2457104.232810& $\pm$0.00012    &s      &R      &SOAO\\
        2457104.290880& $\pm$0.000036   &p      &R      &SOAO\\
        2457130.377900& $\pm$0.000054   &s      &R      &AUKR\\
        2457130.435820& $\pm$0.000057   &p      &R      &AUKR\\
        2457137.322470& $\pm$0.000022   &p      &R      &TUG\\
        2457158.448580& $\pm$0.000064   &p      &R      &AUKR\\
        2457169.303600& $\pm$0.000044   &p      &R      &AUKR\\
        2457422.468150& $\pm$0.000095   &p      &R      &AUKR\\
        2457422.584870& $\pm$0.000078   &p      &R      &AUKR\\
        2457426.553410& $\pm$0.000064   &p      &R      &AUKR\\
        2457426.611750& $\pm$0.000094   &s      &R      &AUKR\\
        2457440.442900& $\pm$0.00016    &p      &R      &AUKR\\
        2457440.501550& $\pm$0.00014    &s      &R      &AUKR\\
        2457440.559700& $\pm$0.00008    &p      &R      &AUKR\\
        2457441.493442& $\pm$0.000022   &p      &R      &AUKR\\
        2457441.551859& $\pm$0.000069   &s      &R      &AUKR\\
        2457441.610196& $\pm$0.000022   &p      &R      &AUKR\\
        2457457.484029& $\pm$0.000034   &p      &V      &AUKR\\
        2457457.542536& $\pm$0.000075   &s      &V      &AUKR\\
        2457457.600768& $\pm$0.000045   &p      &V      &AUKR\\
        2457469.097705& $\pm$0.00023    &s      &R      &SOAO\\
        2457469.155938& $\pm$0.00019    &p      &R      &SOAO\\
        2457469.214385& $\pm$0.00012    &s      &R      &SOAO\\
        2457469.272720& $\pm$0.000048   &p      &R      &SOAO\\
        2457469.331065& $\pm$0.00023    &s      &R      &SOAO\\
        2457480.069250& $\pm$0.000088   &s      &R      &SOAO\\
        2457480.185980& $\pm$0.00017    &s      &R      &SOAO\\
        2457480.244330& $\pm$0.000094   &p      &R      &SOAO\\
        2457480.302610& $\pm$0.00010    &s      &R      &SOAO\\
        2457480.361057& $\pm$0.000077   &p      &V      &AUKR\\
        2457480.419463& $\pm$0.00010    &s      &I      &AUKR\\
        2457492.091380& $\pm$0.00002    &s      &R      &SOAO\\
        2457492.149730& $\pm$0.000019   &p      &R      &SOAO\\
        2457506.331194& $\pm$0.000083   &s      &I      &AUKR\\
        2457506.389490& $\pm$0.000011   &p      &I      &AUKR\\
        2457506.447964& $\pm$0.000036   &s      &V      &AUKR\\
        2457506.506248& $\pm$0.000013   &p      &V      &AUKR\\
        2457506.856370& $\pm$0.000005   &p      &B      &LOAO\\
        2457507.673350& $\pm$0.000011   &p      &V      &LOAO\\
        2457507.731700& $\pm$0.000038   &s      &V      &LOAO\\
        2457507.790090& $\pm$0.00001    &p      &V      &LOAO\\
        2457507.848350& $\pm$0.000031   &s      &V      &LOAO\\
        2457508.782240& $\pm$0.000021   &s      &R      &LOAO\\
        2457508.840580& $\pm$0.000009   &p      &R      &LOAO\\
        2457787.566715& $\pm$0.000035   &p      &B      &AUKR\\
        2457787.625322& $\pm$0.00022    &s      &B      &AUKR\\
        2457803.381877& $\pm$0.00018    &s      &B      &AUKR\\
        2457803.440585& $\pm$0.000037   &p      &B      &AUKR\\
        2457803.498890& $\pm$0.00017    &s      &B      &AUKR\\
        2457803.557359& $\pm$0.00004    &p      &B      &AUKR\\
        2457803.615831& $\pm$0.000095   &s      &B      &AUKR\\
        2457847.502204& $\pm$0.000069   &s      &R      &AUKR\\
        2457847.560560& $\pm$0.00005    &p      &R      &AUKR\\
        2457911.347827& $\pm$0.00015    &s      &R      &AUKR\\
        2457917.358775& $\pm$0.00005    &p      &R      &AUKR\\
        2458242.305835&$\pm$0.00001     &p      &R      &TUG\\
        2458242.364198&$\pm$0.000032    &p      &R      &TUG\\
        2458567.369609&$\pm$0.000026    &p      &R      &AUKR\\
\end{supertabular}

\clearpage
\onecolumn
\section{Parameter uncertainties for Model 6}\label{appendixchi2probabilities}

        \begin{figure*}[h]
                \centering
                \includegraphics[width=480px]{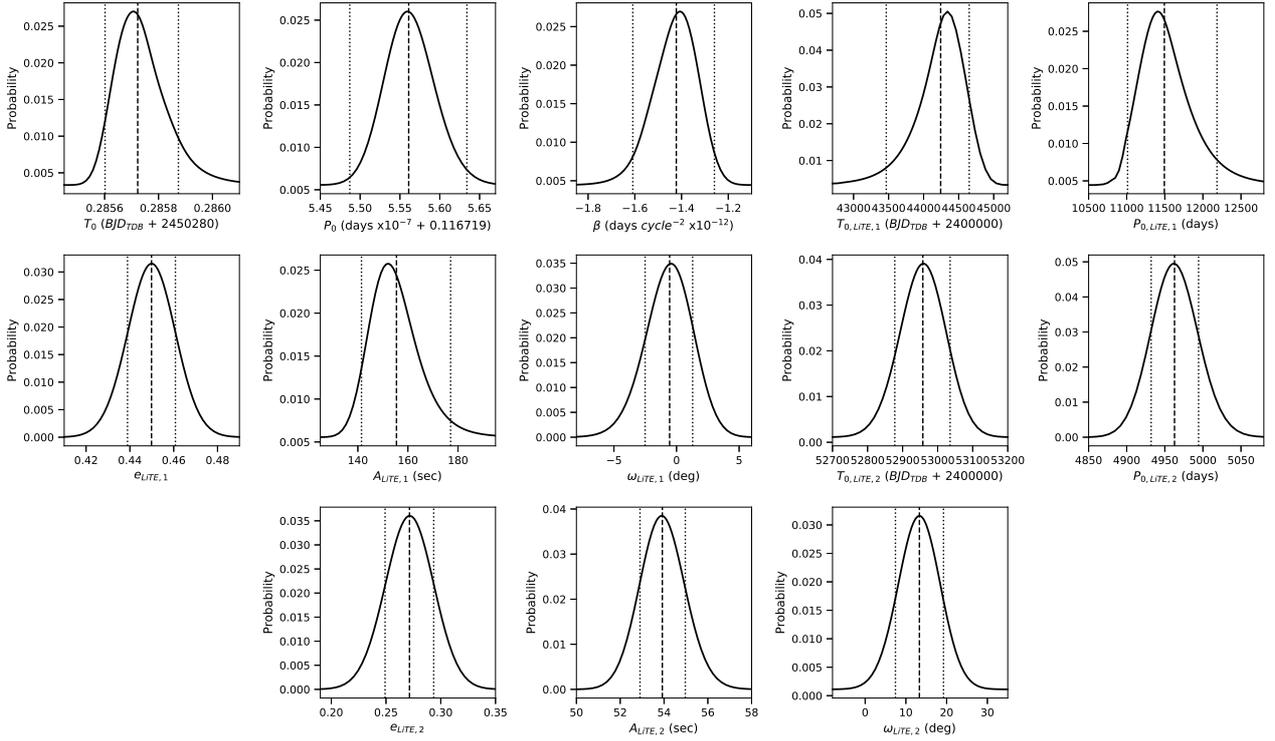}
                \caption{Marginalized probability density distributions ($P(x,y) \propto exp(-\chi^2/2)$) for the parameters of Model 6 as obtained from the $\chi^2$-grid method. The 68.3\% confidence interval and median values are shown with vertical lines (see text for more details).}
                \label{figure:chi2surface_probabilities}
        \end{figure*}

        \begin{figure*}[h]
                \centering
                \includegraphics[width=480px]{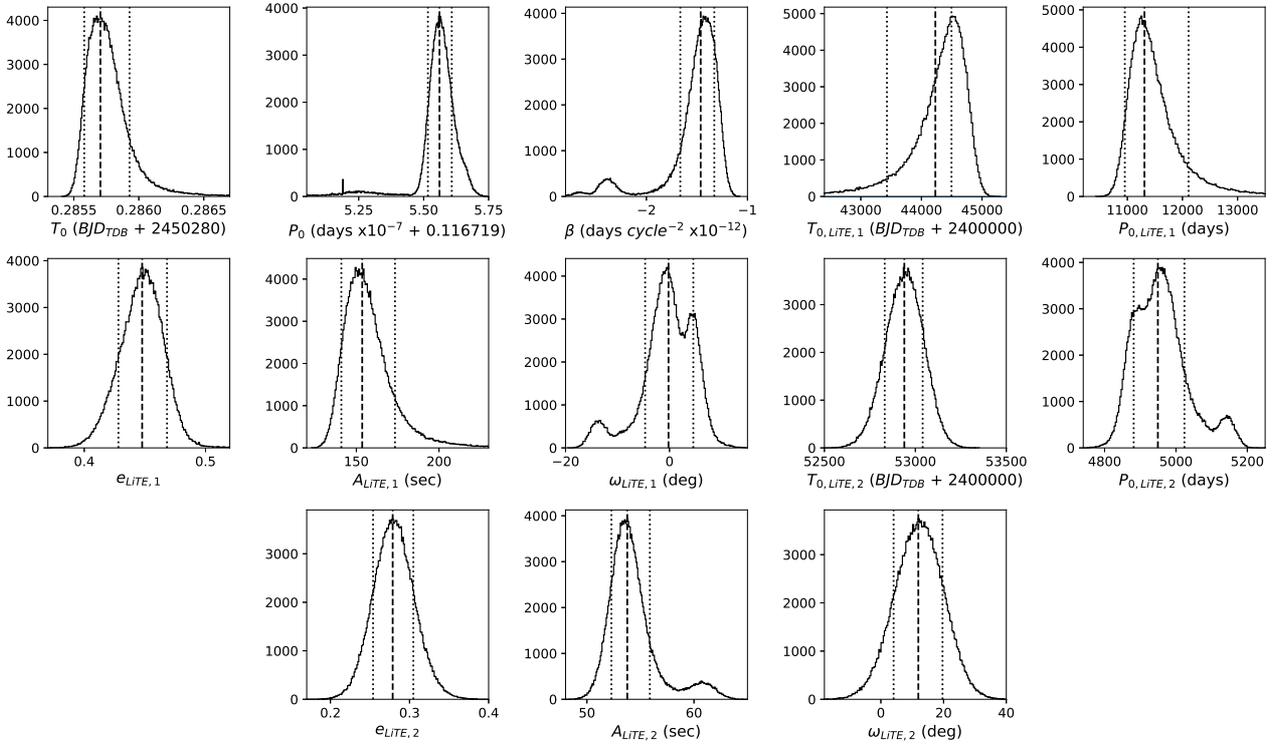}
                \caption{Histograms of parameters from the bootstrap method. $\pm 68.3\%$ confidence limits and median values are shown with vertical lines.}
                \label{figure:bootstrap_histogram}
        \end{figure*}
\clearpage
\section{$\chi^2$ surface grids for Model 6}\label{appendixchi2surface}

\begin{figure*}[h]
                \centering
                \includegraphics[width=500px]{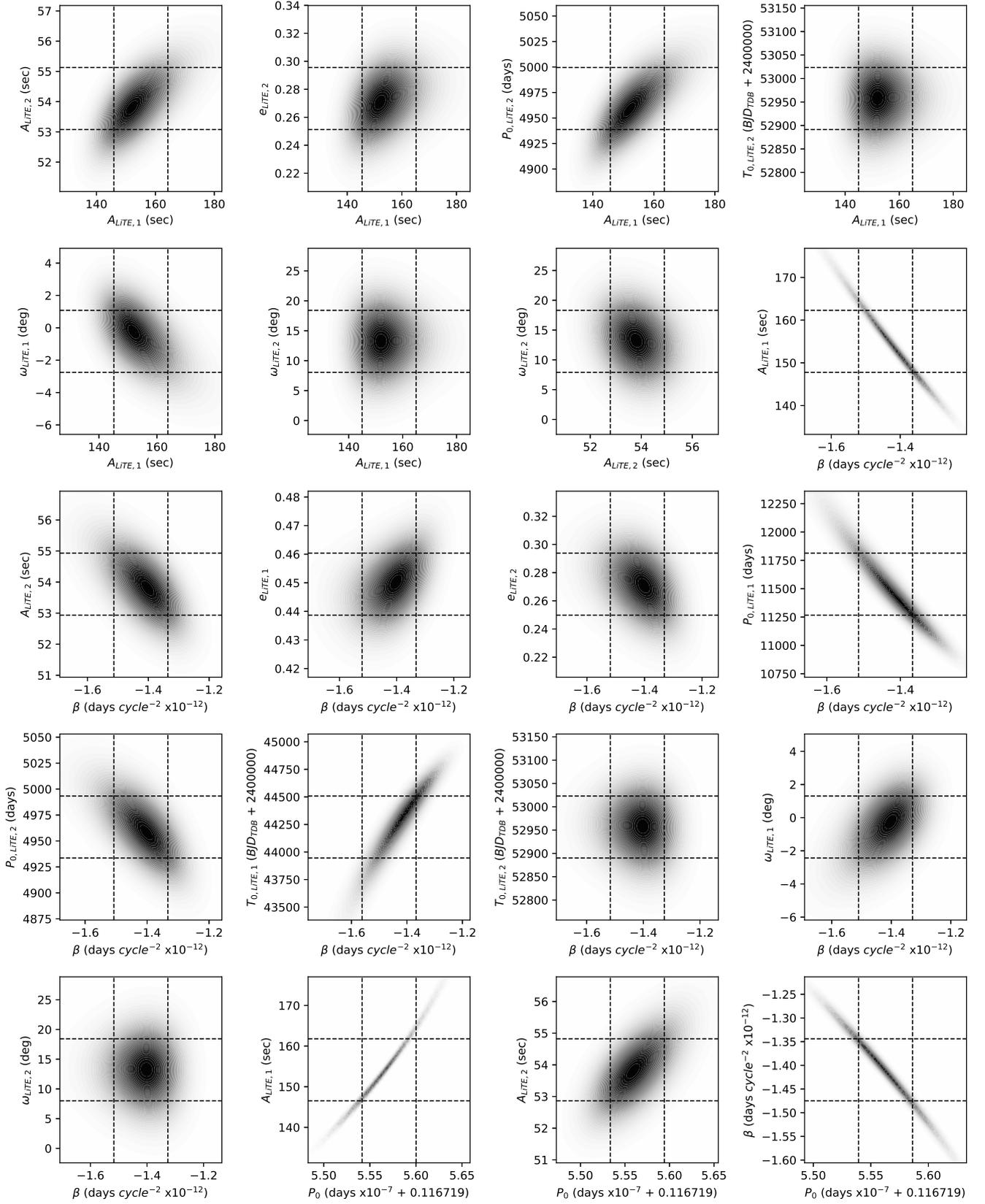}
                \caption{$\chi^2$ surface plots for the pairs of parameters of Model 6. The vertical and horizontal dashed lines represent $\pm 1 \sigma$ values.}
\end{figure*}


\begin{figure*}
        \centering
        \includegraphics[width=500px]{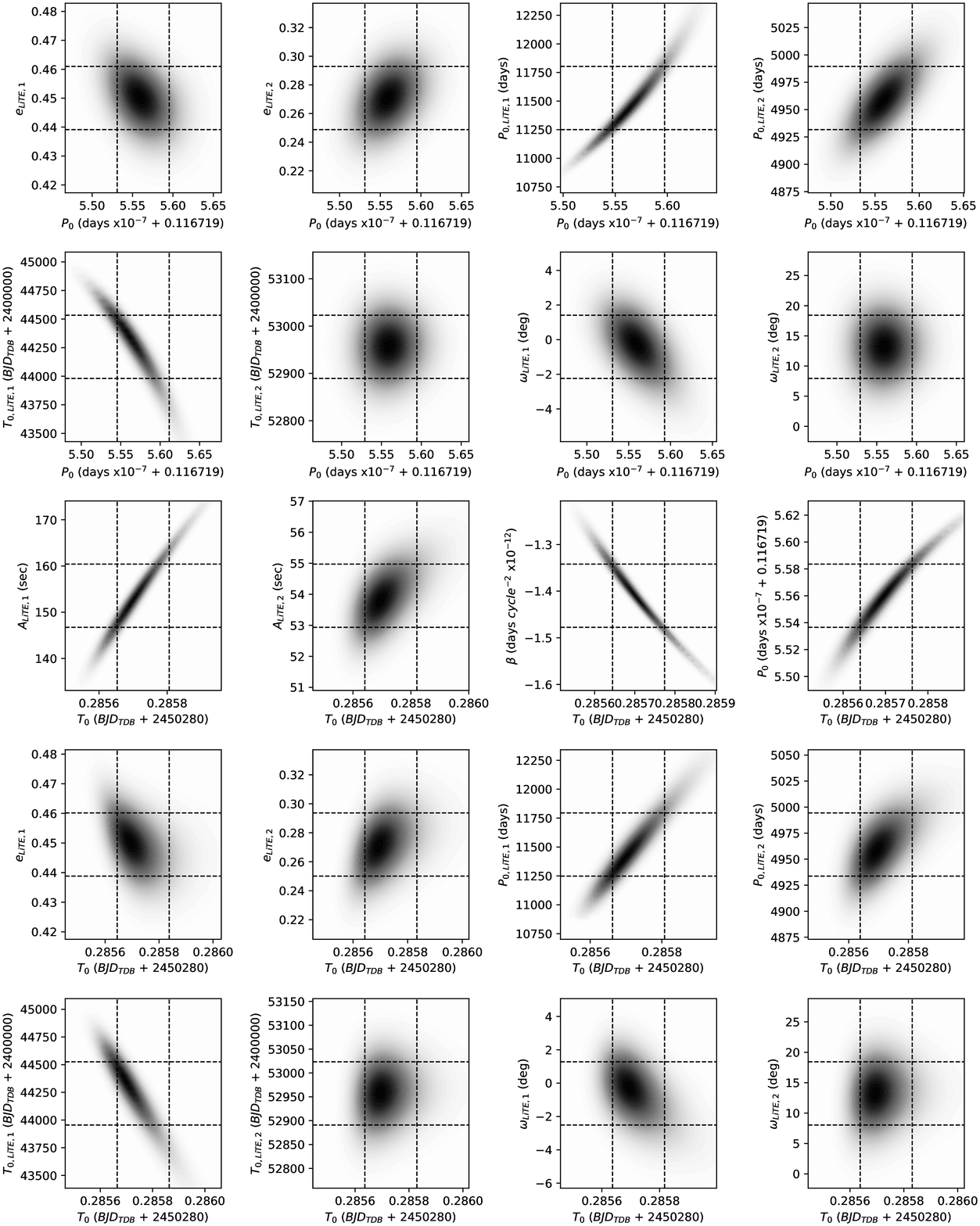}
        \caption{(continued) $\chi^2$ surface plots for the pairs of parameters of Model 6.}
\end{figure*}

\clearpage

\begin{figure*}
        \centering
        \includegraphics[width=500px]{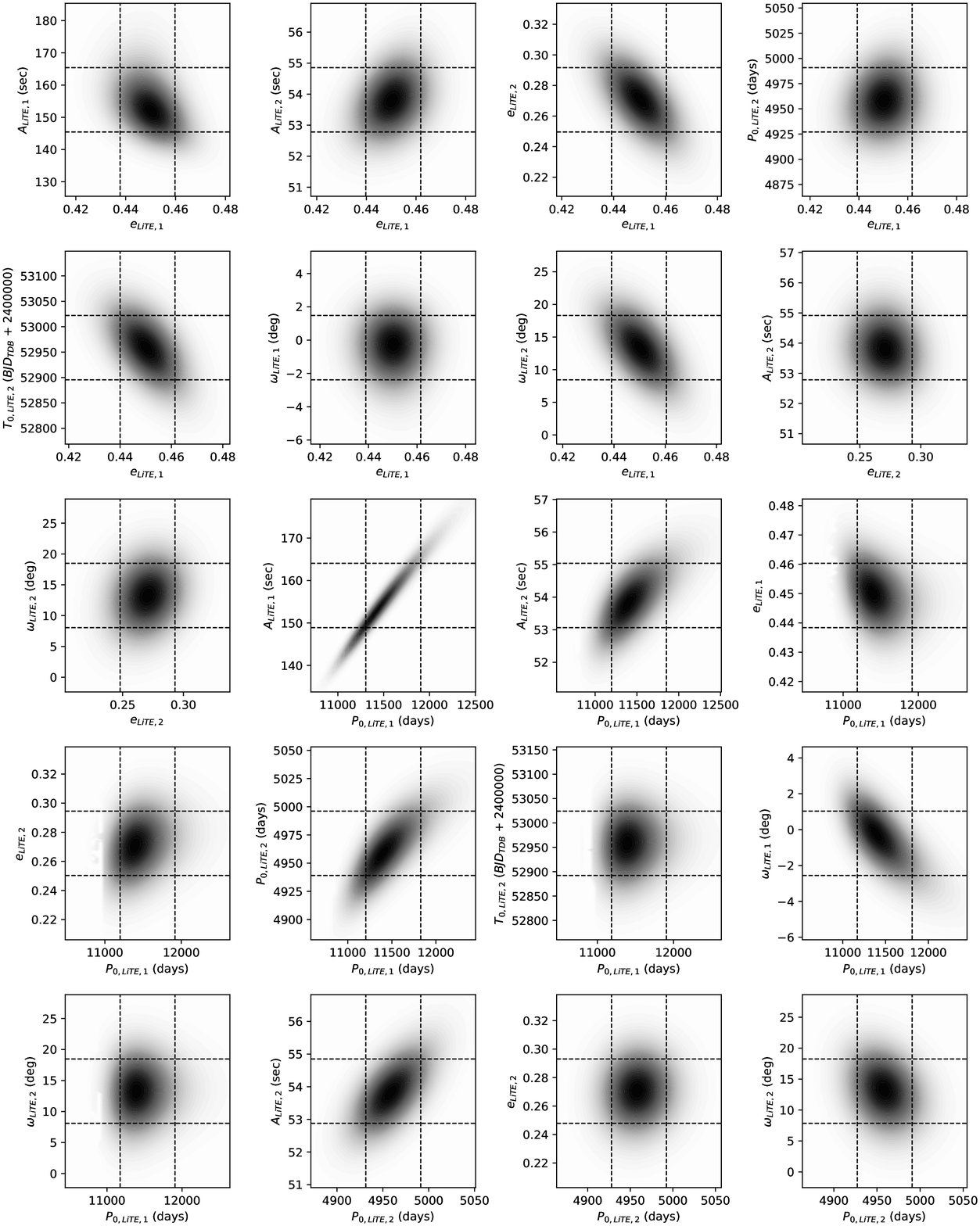}
        \caption{(continued) $\chi^2$ surface plots for the pairs of parameters of Model 6.}
\end{figure*}
\begin{figure*}
        \centering
        \includegraphics[width=500px]{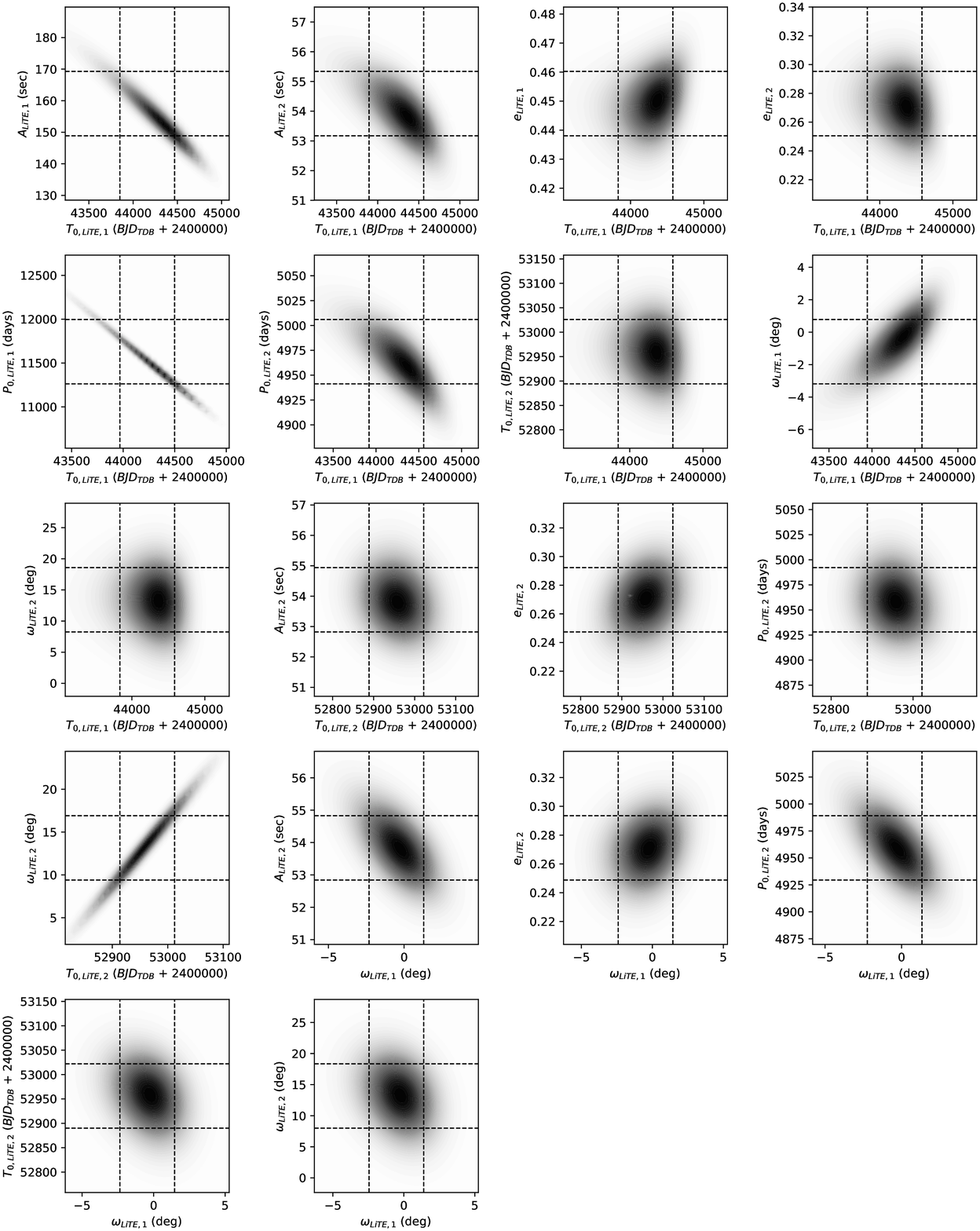}
        \caption{(continued) $\chi^2$ surface plots for the pairs of parameters of Model 6.}
\end{figure*}
%

\end{appendix}

\end{document}